 \documentclass[iop,appendixfloats]{emulateapj}

\newcommand{\bit}[1]{ \textbf{\emph{#1}} }

\newcommand{\Mr}{M_{\rm R}}

\newcommand{\etatens}{\overline{\overline{\eta}}}

\newcommand{\Grav}{\Phi_{\rm g}}

\newcommand{\Cs}{C_{\rm s}}
\newcommand{\CsT}{C_{\rm s}^{\rm T}}

\newcommand{\Vtot}{\bit{V}}

\newcommand{\Vp}{V_{\rm P}}
\newcommand{\Va}{V_{\rm A}}
\newcommand{\Vr}{V_{R}}
\newcommand{\Vth}{V_{\rm \theta}}
\newcommand{\Vphi}{V_{\rm \phi}}
\newcommand{\Vk}{V_{\rm K}}

\newcommand{\Btot}{\bit{B}}
\newcommand{\Bp}{B_{\rm P}}
\newcommand{\Br}{B_{\rm R}}
\newcommand{\Bth}{B_{\rm \theta}}
\newcommand{\Bphi}{B_{\rm \phi}}

\newcommand{\Jtot}{\bit{J}}

\newcommand{\Jth}{\vec{J}_{\rm \theta}}

\newcommand{\ass}{{\alpha}_{\rm ssm}}
\newcommand{\am}{{\alpha}_{\rm m}}

\newcommand{\Ht}{ H_{\rm T} }

\newcommand{\Fe}{ F_{\rm \eta} }

\newcommand{\MUzero}{ \mu_{\rm 0} }
\newcommand{\MUdisk}{ \mu_{\rm D} }

\newcommand{\MUcrit}{ \mu_{\rm crit} }

\newcommand{\Ep}{ \eta_{\rm P}}
\newcommand{\Et}{ \eta_{\rm T}}

\newcommand{\om} { \Omega_{\rm F} }
\newcommand{\omz} { \Omega_{\rm F,0} }
\newcommand{\ra} { r_{\rm A} }

\newcommand{\vmax} { v_{\rm max} }

\newcommand{\rhod} { \rho_{\rm D} }
\newcommand{\vpd}  { V_{\rm P,D} }
\newcommand{\bpd}  { B_{\rm P,D} }
\newcommand{\rhoa} { \rho_{\rm A} }
\newcommand{\vpa}  { V_{\rm P,A} }
\newcommand{\bpa}  { B_{\rm P,A} }

\newbox\grsign \setbox\grsign=\hbox{$>$}
\newdimen\grdimen \grdimen=\ht\grsign
\newbox\laxbox \newbox\gaxbox
\setbox\gaxbox=\hbox{\raise.5ex\hbox{$>$}\llap
     {\lower.5ex\hbox{$\sim$}}}\ht1=\grdimen\dp1=0pt
\setbox\laxbox=\hbox{\raise.5ex\hbox{$<$}\llap
     {\lower.5ex\hbox{$\sim$}}}\ht2=\grdimen\dp2=0pt
\newcommand{\gax}{$\mathrel{\copy\gaxbox}$}
\newcommand{\lax}{$\mathrel{\copy\laxbox}$}

\shorttitle{Jets from Disks}
\shortauthors{Stepanovs \& Fendt}

\usepackage{ctable} 
\usepackage{epstopdf}
\begin{document}

\title{An extensive numerical survey of the correlation between outflow dynamics and accretion disk magnetization}
\author{Deniss Stepanovs\altaffilmark{1}}
\author{Christian Fendt}
\affil{Max Planck Institute for Astronomy, K\"onigstuhl 17, D-69117 Heidelberg, Germany}
\email{fendt@mpia.de}
\altaffiltext{1}{Fellow of the
                 \textit{International Max Planck Research School for Astronomy and 
                         Cosmic Physics at the University of Heidelberg (IMPRS-HD)
                        }
                }

\begin{abstract}
We investigate the accretion-ejection process of jets from magnetized accretion disks.
We apply a novel approach to the jet-launching problem in order to obtain correlations between the 
physical properties of the jet and the underlying disk.
We extend and confirm the previous works of \citet{2009MNRAS.400..820T} and \citet{2010A&A...512A..82M}
by scanning 
a large parameter range for the disk magnetization, $\mu_{\rm D} = 10^{-3.5} ... 10^{-0.7}$.
We disentangle the disk magnetization at the foot point of the outflow as the main parameter that 
governs the properties of the outflow.
We show how the four jet integrals known from steady-state MHD are correlated to the disk magnetization 
at the jet foot point.
This agrees with the usual findings of the steady-state theory, however, here we obtain these 
correlations from time-dependent simulations that include the dynamical evolution of the disk 
in the treatment.
In particular, we obtain robust correlations between the local disk magnetization and 
(i)the outflow velocity, 
(ii) the jet mass loading, 
(iii) jet angular momentum, and 
(iv) the local mass accretion rate.
Essentially we find that strongly magnetized disks launch more energetic and faster jets, and, 
due to a larger 
Alfv\'en lever arm, these jets extract more angular momentum from the underlying disk.
These kind of disk-jet systems have, however, a smaller mass loading parameter and a lower mass 
ejection-to-accretion ratio.
The jets are launched at the disk surface where the magnetization is $\mu(r,z) \simeq 0.1$.
The magnetization rapidly increases vertically providing the energy reservoir for subsequent 
jet acceleration.
We find indication for a critical disk magnetization $\mu_{\rm D} \simeq 0.01$ that separates the
regimes of magnetocentrifugally-driven and magnetic pressure-driven jets.
\end{abstract}

\section{Introduction}
%
%
Astrophysical jets consist of magnetized material moving as highly collimated beams with high 
velocity.
These outflows are typically bipolar in nature and are an ubiquitous phenomenon in a variety of
astrophysical sources.
Today it seems to be commonly accepted that jets and outflows are launched from a magnetized accretion
disk surrounding a central object
\citep{2007prpl.conf..277P, 2007prpl.conf..231R, 2015SSRv..191..441H}.

A good number of studies have been devoted to the understanding of the launching process, 
i.e. the {\em transition from accretion to ejection}. 
These studies consider steady-state MHD 
(see e.g. \citealt{1993ApJ...410..218W, 1995ApJ...444..848L, 1995A&A...295..807F,
2010MNRAS.401..479K, 2011MNRAS.412.1162S})
or time-dependent numerical simulations (see e.g.
\citealt{1998ApJ...508..186K,2002ApJ...581..988C, 2004ApJ...601...90C,  
2006A&A...460....1M, 2007A&A...469..811Z, 2009MNRAS.400..820T, 2010A&A...512A..82M, 2012ApJ...757...65S, 
2013ApJ...774...12F}).
Recent works consider the accretion-ejection evolution for long time scales and
on a large computational grid \citep{2014ApJ...793...31S} and jet-launching by a large-scale
magnetic field self-generated by a mean-field disk dynamo \citep{2014ApJ...796...29S},
and also the non-axisymmetric launching process of disk jets \citep{2015ApJ...814..113S}.
However, despite all the efforts over decades, the fundamental question of what kind of disks
drive jets and what kind of disks do not, is still not answered.

An essential ingredient for jet launching is a strong poloidal magnetic field,
as shown by seminal analytical papers 
\citep{1982MNRAS.199..883B,1983ApJ...274..677P,1984PASJ...36..105U,1997A&A...319..340F},
and further confirmed by numerical simulations 
\citep{1985PASJ...37..515U,1995ApJ...439L..39U,1997ApJ...482..712O,1998ApJ...508..186K,2002ApJ...565.1035K,
2002ApJ...581..988C,2002A&A...395.1045F,2007A&A...469..811Z}.

How the accretion disk magnetization affects the jet launching has been studied in detail by
\citet{2009MNRAS.400..820T} and \citet{2010A&A...512A..82M}.
In \citet{2009MNRAS.400..820T} no disk viscosity was included and angular momentum transfer
was purely magnetic. 
These simulations confirmed earlier analytical works (e.g. \citealt{1995A&A...295..807F})
suggesting that only a large - near equipartition - magnetization is able to produce
steady-state super fast magneto sonic jets.
The simulations of \citet{2010A&A...512A..82M} included disk viscosity and thus also viscous
angular momentum transport.
These authors showed - surprisingly - that also a weak disk magnetization could launch
super-fast jets. 
The results were interpreted in the framework of {\em warm} disk wind solutions 
\citep{2000A&A...353.1115C}, suggesting that numerical mass diffusion may exist that could
mimic heating and physical mass diffusion.
The acceleration of the material loaded onto the jet is achieved in the usual Blandford-Payne
magneto-centrifugal as long as the jet magnetization (the ratio of Poynting flux to kinetic 
energy flux) is larger than unity. 
In this case, also super-fast jet velocities could be obtained.
The authors derived two main criteria for jet launching from their study. 
One is that the disk magnetization cannot be too small, as the jet magnetization is proportional 
to the disk magnetization.
The other is that the diffusive process of mass loading that brings mass from the zone of 
resistive MHD (disk) to the zone of ideal MHD (jet) needs to be efficient.

It was further suggested that super fast magnetosonic jets can only be launched from disk 
surface areas where the magnetization $\mu \sim B_P^2 / P$ becomes about unity.
That area seems to be concentrated to about 5 inner disk radii.

\citet{2012ApJ...757...65S} and \citet{2013ApJ...774...12F}
report a variation of the disk Alfv\'en speed by a factor ten during their simulations.
Usually the disk magnetic diffusivity prescription is parameterized by the Alfv\'en speed. 
Therefore, besides the magnetization, also the disk diffusivity is substantially modified during 
the simulations - a fact that severely affects the disk evolution and also the mass loading 
from the disk into the outflow.

\citet{2010A&A...512A..82M} and \citet{2012ApJ...757...65S} find outflows also from weakly 
magnetized disks.
\citet{2012ApJ...757...65S} conclude that the driving of these outflows was not in the regime of
the magneto-centrifugal Blandford-Payne mechanism, but in the regime of a magnetic pressure-driven 
tower outflow.
In \citet{2014ApJ...793...31S} (hereafter paper I), we have demonstrated in more detail that it 
is 
essential to relate the {\em actual}\footnote{With {``}actual{''} we denote the conditions at a 
certain time, thus snapshots in time of the set of physical parameters.} 
jet  properties to the {\em actual} disk properties, as both disk and jet evolve dynamically in time.
Thus, the initial simulation parameters such as the initial magnetization, do not necessarily 
characterize the jet evolution, as these parameters are substantially changing during the disk 
evolution.

In the present paper, we continue to investigate the correlation between disk magnetization and
jet launching by applying a new approach based on the long-term evolution of the disk-jet system.
We take advantage from the fact that due to the loss of disk mass by accretion and ejection, the 
disk physics slowly changes in time.
However, each time interval of the evolution can be considered as a quasi steady state.
By comparing the disk and jet parameters at {\em different times} - thus {\em snapshots} of the 
accretion-ejection structure at different times - we effectively compare the jet launching
conditions of {\em different disks}.

Our new simulations extend those presented in paper I, allowing us to probe a broad parameter space 
concerning the disk magnetization.
As a result, we will present numerical correlations between the disk magnetization and typical outflow
parameters over a wide range of disk magnetization.

%
In Section~2 we introduce to our general model approach, in particular our approach to analyze the slowly
evolving the disk-jet system.
In Section~3 we present our model setup, the choice of parameters, and discuss the general evolution of 
the system (summarizing paper I in this point).
In Section~4 we discuss physical properties of the jet by introducing the jet steady-state MHD integrals
and show how to derive them from our numerical simulations.
In Section~5 we connect the disk magnetization to various jet properties and
present general correlations between them. 
In Section~6 we discuss the relevance for the jet ejection mechanism and compare our results
to other studies, mainly to the paper of \citet{2010A&A...512A..82M}.
We summarize our results in Section~7.

\section{General model approach}
Before we detail our model setup, we summarize our approach of addressing the jet-launching problem 
for a range of disk magnetizations, thereby mentioning several key points.
The novelty of our approach lies in the fact that {\em we take advantage of the long-term disk 
evolution}, 
mainly triggered by the disk mass loss during the simulation.
This enables us to compare {\em different disks} of different magnetization.

We perform {\em large-scale simulations} which ignores the microphysics of the accretion disk.
This approach allows us to evolve both the accretion disk and the jet extending far 
from the disk in the same computational domain.
The drawback is that the peculiar realm of the disk turbulence enters the equations only as an 
effective profile of the magnetic diffusivity.

We further note that the magnetic diffusivity profile follows the standard prescription 
used in the literature \citep{2007A&A...469..811Z,2010A&A...512A..82M}, thus assuming 
a simple parameterization of the magnetic diffusivity strength by the diffusivity parameter 
$\am$. 
However, we keep the diffusivity scale height constant in time.

We summarize the essential of our approach in linking the disk quantities to the outflow 
quantities as follows:
(1) We start a simulation with a certain initial disk magnetization and strength of 
          the magnetic diffusivity.\\
(2) As the accretion-ejection structure evolves in time a quasi-steady state
        is reached at which advection and diffusion balance.\\
(3) Since the disk slowly looses mass to the central object and into the outflow, the
          disk magnetization slowly changes in time. This time scale can be estimated to
          10,000 time units.
          We decided to show the simulation results at 10,000 time units for the purpose 
          of this paper, although most simulations run longer.\\
(4) As a consequence also the outflow characteristics change slowly in time.\\
(5) We continuously measure the disk magnetization and the outflow parameters, both
          suitably averaged at certain radius and altitude along the outflow.
          This provides us with a time series of the actual magnetization and
          actual outflow parameters at a certain time. \\
(6) Combining the actual physical properties for a certain time, we can then
          provide correlations between these physical properties - in particular, we will
          relate the outflow physical properties to the disk magnetization. \\
(7) We may then interrelate the outflow parameters that result from the simulation to the 
          asymptotic jet parameters by using the steady-state MHD integrals.
          This is possible since the slow disk evolution in quasi-steady state that takes
          much longer than the outflow propagation times scale that is the Keplerian 
          time scale at the foot point of the outflow.

Since the disk magnetization and also the outflow parameters vary in time, we obtain a general correlation 
between the disk magnetization and the outflow properties.
In principle, such correlations can be obtained for a range of magnetic field foot point radii along the disk.
However, as the evolution of the outer disk proceeds considerably slower, it is rather difficult to probe 
a broad range of magnetization at these radii.

\section{Model setup}
We apply the MHD code PLUTO \citep{2007ApJS..170..228M}, version 4.0,
solving the time-dependent, resistive MHD equations on a spherical grid $(R,\theta)$. 
We refer to $(r,z)$ as cylindrical coordinates.

The code numerically solves for the mass conservation,
\begin{equation}
\frac{\partial\rho}{\partial t} + \nabla \cdot(\rho \Vtot)=0,
\end{equation}
with the plasma density $\rho$ and flow velocity $\Vtot$,
the momentum conservation,
\begin{equation}
\frac{\partial \rho \Vtot}{\partial t} + 
\nabla \cdot \left[ \rho  \Vtot \Vtot +
\left(P + \frac{\Btot \cdot \Btot}{2}\right)I - 
\Btot\Btot \right] +\rho \nabla \Grav = 0
\end{equation}
with the thermal pressure $P$ and the magnetic field $\Btot$. 
We consider a central object of point mass $M$ with the gravitational potential 
$\Grav = - G M/R$ with the gravitational constant $G$.
Note that equations are written in non-dimensional form, and as usual the
factor $4\pi$ is neglected for the magnetic field.
We apply a polytropic equation of state, $P \propto \rho^{\gamma}$, with $\gamma=5/3$.

The code further solves for the conservation of energy,
\begin{eqnarray}
\frac{\partial e}{\partial t} & + &  \nabla \cdot \left[ \left( e + P + \frac{\Btot \cdot \Btot}{2} \right) \Vtot
                              -  (\Vtot \cdot \Btot)\Btot + \etatens \Jtot \times \Btot \right]  \nonumber \\
                              & = & -\Lambda_{\rm{cool}},
\end{eqnarray}
with the total energy density,
\begin{equation}
e = \frac{P}{\gamma -1} + 
    \frac{\rho \Vtot \cdot \Vtot}{2} + 
    \frac{\Btot \cdot \Btot}{2} + \rho \Grav,
\end{equation}
given by the sum of thermal, kinetic, magnetic, and gravitational energy,
respectively. 
The electric current density is denoted by $\Jtot=\nabla \times \Btot$.
As shown by \citet{2013MNRAS.428.3151T}, cooling may indeed play a role for jet launching, 
influencing both jet density and velocity.
For the sake of simplicity we set the cooling term equal to Ohmic heating,
$\Lambda_{\rm{cool}} = -\etatens \Jtot \cdot \Jtot$.
Thus all generated heat is instantly radiated away.

The magnetic field evolution is governed by the induction equation
\begin{equation}
\frac{\partial \Btot}{\partial t} = 
\nabla\times (\Vtot \times \Btot  -  \etatens \Jtot),
\end{equation}
In general the magnetic diffusivity is defined as a tensor, $\etatens$.
We assume the diffusivity tensor to be diagonal with the non-zero components
$\eta_{\rm \phi\phi} \equiv  \eta_{\rm p}$, and $\eta_{\rm rr}=\eta_{\rm zz} \equiv  \eta_{\phi}$,
where we denote $\eta_{\rm p}$ as the {\em poloidal magnetic diffusivity},
and $\eta_{\phi}$ as the toroidal magnetic diffusivity, respectively 
(see Sect. 3.2. and \citealt{2012ApJ...757...65S} and \citealt{2013ApJ...774...12F}).

We normalize the system of equation as follows.
Lengths are given in units of the inner disk radius $R_0$.
Velocities are given in units of $V_{\rm K,0}$, corresponding to the Keplerian speed at the inner radius.
The time unit is $T_0 = R_0 / V_{\rm K,0}$, which we call the {\em dynamical time step}.
The time $2\pi T_0 = 2\pi R_0 /V_{\rm K,0} $ corresponds to one revolution at the inner disk radius.
Densities are given in units of $\rho_0$, the disk density at the inner radius at the disk midplane.
For further details, in particular the astrophysical scaling, we refer to paper I.

We apply a numerical grid with equidistant spacing in $\theta$-direction,
but stretched cell sizes in radial direction, considering $\Delta R = R \Delta\theta$.
Our computational domain of a size $R=(1, 1500 R_0),\theta=(0,\pi/2)$ is discretized with $(N_R \times N_\theta)$ grid cells.
We use a general resolution of $N_\theta = 128$.
In order to cover a factor 1500 in radius, we apply $N_R = 600$.
This gives a resolution of 16 cells per disk height ($2 \epsilon$) in the general case.
However, we have also performed a resolution study applying a resolution twice high (or low, respectively),
thus using $256 \times 1200$ (or $64 \times 300$) cells for the whole domain, or 35 (9) cells per disk height
(see paper I).

We select the Harten-Lax-van Leer (HLL) Riemann solver together with a third-order order 
Runge-Kutta scheme for time evolution and the PPM (piecewise parabolic method) reconstruction 
of \citep{1984JCoPh..54..174C} for spatial integration.
The magnetic field evolution follows the method of Constrained Transport \citep{2004JCoPh.195...17L}.

\subsection{Initial conditions}
For the initial conditions we follow a standard setup, commonly used in the literature 
\citep{2007A&A...469..811Z, 2012ApJ...757...65S, 2013ApJ...774...12F},
and further detailed in paper I.
The initial structure of the accretion disk is calculated from the steady state force equilibrium,
\begin{equation}
\nabla P  +\rho \nabla \Grav  
- \Jtot \times \Btot - \frac{1}{R} \rho \Vphi^2 
 ({\bit e}_R \sin\theta + {\bit e}_\theta \cos\theta) = 0.
\end{equation}
We solve this equations assuming radial self-similarity (see paper I).

An essential non-dimensional parameter governing the initial disk structure is 
the ratio $\epsilon$ between the isothermal sound speed $\CsT = \sqrt{P/\rho}$ and
the Keplerian velocity $\Vk = \sqrt{GM/r}$, evaluated at the disk midplane,
$ \epsilon \equiv \left[ \CsT / \Vk \right]_{\,\theta=\pi/2}. $
This quantity determines the disk thermal scale height $\Ht = \epsilon r$. 
We assume $\epsilon = 0.1$ initially\footnote{
Note that for the rest of the paper, when discussing the dynamical properties of disk 
and outflow, we consider the {\em adiabatic sound speed} $\Cs=\sqrt{\gamma P/\rho}$.}
The thermal disk height $\Ht$ will change during the disk evolution, however,
it is not further used as a parameter.
In contrast, we define a geometrical disk height, namely the scale length where disk
density and rotation significantly decrease, $H \equiv 2 \epsilon r$.
This parameter remains fixed in time, and is used to define the initial condition
and the vertical profile of the magnetic diffusivity (see Sect.~3.3).

Following \citet{2007A&A...469..811Z}, our reference simulation is initialized with 
a poloidal magnetic field only.
This is defined by the 3-component of the vector potential with $\Btot = \nabla \times A \bit{e}_{\phi}$, 
\begin{equation}\label{eq:Bpot}
A = \frac{4}{3} B_{\rm p,0} R^{-1/4} \frac{m^{5/4}}{ (m^2 + ctg^2\theta)^{5/8} }\,\,.
\end{equation}
The parameter $B_{\rm p,0}$ determines the strength
of the initial magnetic field.
The parameter $m$ determines the degree of curvature of the poloidal magnetic field 
lines.
As we have shown in paper I, the long-term evolution of the disk-jet structure
is insensitive to this parameter, 
since due to advection and diffusion processes in the disk and the existence of
an outflow, the magnetic field profile is changed substantially over time.
We therefore assume $m=0.5$ in general.

The relative strength of the magnetic field is governed by the magnetization parameter,
generally defined as $\mu_{\rm D} = [ \Bp^2/2P ]_{\,\theta=(\pi/2)}$,
the ratio between the poloidal magnetic field pressure and the thermal pressure.
As it has been shown \citep{2010A&A...512A..82M, 2012ApJ...757...65S, 2014ApJ...793...31S}
the magnetic field distribution substantially changes over time.
Therefore, the disk-jet dynamics is governed by the {\em actual} disk magnetization.
The {\em initial} disk magnetization is denoted by $\MUzero = 0.5 \epsilon / B_{\rm p,0}$
and is calculated at the inner disk radius at the midplane and is set to be constant 
with radius.
Typically, we apply $\MUzero \approx 0.01$. 
We will further apply the notation $\MUdisk = \MUdisk(R,t)$ for the time-dependent disk 
magnetization measured at the midplane. 
This will be the leading parameter of our paper.

As mentioned above, we obtain the initial disk density $\rho_{\rm disk} (R,\theta)$ and
disk pressure $P_{\rm disk}(R,\theta)$ from integrating equation (6). 
As for $\MUdisk$, the $\rho_{\rm D}(R)$ denotes the disk density along the midplane.
Outside the disk the gas and pressure distribution is defined as hydrostatic {``}corona{''},
\begin{equation}
\rho_{\rm cor} = \rho_{\rm cor,0} R^{-1/(\gamma - 1) } ,
\,\, 
P_{\rm cor} = \frac{\gamma -1}{\gamma} \rho_{\rm cor,0}  R^{-\gamma /(\gamma - 1)},
\end{equation}
where $\rho_{\rm cor,0} = \delta \rho_{\rm disk} (R=1,\theta=\pi/2)$ with $\delta = 10^{-3}$
and the index 0 again referring to the initial value at the inner disk radius.

Although it is common to define an initial accretion velocity that balances the imposed diffusivity 
$V_R = \eta J_\phi /B_\theta$, 
we find that with our parameter setup this is not necessary (see paper I). 

\subsection{Boundary conditions}
We apply the same boundary conditions as described in paper I.
These are the standard symmetry conditions along the {\em rotational axis} and the 
{\em equatorial plane}.
Along the radial boundaries of the domain, we distinguish two different areas.
That is (i) a {\em disk boundary} for 
$\theta > \frac{\pi}{2}-2\epsilon$\footnote{Note that $2\epsilon \approx \arctan (2\epsilon)$}, 
and (ii) a {\em coronal boundary} for $\theta<\frac{\pi}{2}-2\epsilon$,
and consider different conditions along them.

Along the {\em inner radial boundary} for all simulations we impose 
a constant slope for the poloidal component of the magnetic field.
The magnetic field direction is axial near the axis, $\theta=0$, while at the inner disk radius the
inclination is $70^\circ$ with respect to the disk surface. 
The method of constraint transport requires the definition of only the tangential component, thus to
prescribe $\Bth$ along the innermost boundary.
The normal component $\Br$ follows from solving $\nabla \cdot \Btot =0$.
In order to implement the prescription of a constant magnetic field angle, we solve $\nabla \cdot \Btot =0$, 
taking into account the ratio of the cell-centered magnetic field components $\Bth / \Br = - \tan(\varphi)$. 
We start the integration from the axis ($\theta = 0$), where $\Bth = 0$. 
Thus, by fixing the slope of the magnetic lines, we allow the magnetic field strength to vary.

Along the {\em inner coronal boundary}, we prescribe a weak inflow into the domain with $\Vp = 0.2$. 
This is meant to stabilize the inner coronal region between the rotational axis and the disk jet.
By varying the slope of the magnetic field along this inner corona in the range of 
60 - 80 degrees w.r.t. midplane, we found that it only slightly affects the slope of the innermost magnetic field 
lines. 
The global structure of the magnetic field is instead mainly governed by the diffusivity prescription.
Since the inner boundary by design models the magnetic barrier of the star, we choose a rather steep
slope in order to avoid the disk magnetic flux entering the coronal region. 

Across the {\em inner disk boundary} (that is the accretion boundary) we extrapolate both the density and pressure
by power laws, $\rho R^{3/2} = const$, and $P R^{5/2} = const$, respectively.
Both the toroidal magnetic field as well as the toroidal velocity components are set to vanish at the 
inner coronal boundary, $\Bphi = 0$, $\Vphi = 0$.
For the inner disk boundary, we further apply the condition $\Bphi \sim 1/r$ ($J_\theta = 0$), and extrapolate 
the radial and the toroidal velocity by power laws, 
$\Vr R^{1/2} = const$, and $\Vphi R^{1/2} = const$,
respectively, while $\Vth = 0$.
For the inner disk boundary, only negative radial velocities are allowed, making the boundary to 
behave as a {\em sink}, that absorbs all material that is delivered by the accretion disk at the 
inner disk radius.

The {\em outer boundary conditions} have only little influence on the evolution of the jet launched from the 
very inner disk,
as the application of spherical coordinates provides an opportunity to use a much larger simulation 
domain compared to cylindrical coordinates.
We therefore extrapolate $\rho$ and $P$ with the initial power laws and apply the standard PLUTO 
outflow conditions for $\Vr, \Vth, \Vphi$ at the outer boundary, thus zero gradient conditions.
We further require $\Bphi \sim 1/r$ ($\Jth = 0$) for the toroidal magnetic field component,
while a simple outflow condition is set for $\Bth$\footnote{Note that this condition may imply a 
Lorentz force that is confining along the outer boundary, as discussed by \citet{1999ApJ...516..221U}.
This Lorentz force may affect the collimation and acceleration of the outflow. However, our
outer boundary is at such a large distance that we consider these effects, if present, as marginal.
In addition, a spherical grid minimizes this potential effect \citep{1999ApJ...516..221U}.}.
Again $\Br$ is obtained from the $\nabla \cdot \Btot =0$.

For the radial velocity component we distinguish between the coronal region, where we require positive
velocities $\Vr \geq 0$, and the disk region, where we enforce negative velocities $\Vr \leq 0$.

\subsection{The prescription for the magnetic diffusivity}
The prescription for the magnetic diffusivity is essential for the evolution of the disk dynamics. 
Here we apply the same - so-called standard - prescription that is described in detail in paper I. 
In the following, we repeat some essentials from paper I.

\subsubsection{General approach}
The magnetic diffusivity in accretion disks is considered to be of turbulent origin.
Magnetized disks are subject to the magneto-rotational instability (MRI) for moderate field strengths 
\citep{1991ApJ...376..214B, 2013EAS....62...95F}, 
and to the Parker instability \citep{2010MNRAS.405...41G, 2008A&A...490..501J} for stronger disk 
magnetization.
For strong fields the MRI modes become suppressed \citep{2013EAS....62...95F}. 
On the other hand, a strong magnetic field may become buoyant, leading to the Parker instability.
While the MRI is confined within the disk, the Parker instability operates closer to the surface 
of the disk where the toroidal magnetic field is stronger.

How the Parker instability operates in accretion disk is not quite obvious.
\citet{1994A&A...287..297F} have found that under certain conditions a turn-over may happen that
re-directs the up-lifted material back to the disk midplane.

A self-consistent study of the origin of the turbulence is beyond the scope of our paper.
We therefore need specific prescription of the magnetic diffusivity.
We apply an $\alpha$-prescription \citep{1973A&A....24..337S} for the magnetic diffusivity, implicitly 
assuming that the diffusivity has a turbulent origin.
In general the diffusivity distribution has three major parameters.
That are the radial profile, the vertical profile, and the strength of diffusivity.
The vertical profile may extend up to a level above the pressure scale height of the disk
(see \citealt{2010MNRAS.405...41G, 2012ApJ...757...65S, 2013ApJ...774...12F}).
The radial profile is essential for the advection of disk material across the magnetic field
penetrating the disk. 
Only for high magnetic diffusivity in the outer disk sufficient mass supply from the outer to the 
inner disk is possible, from where disk material is advected towards the central object or is
ejected into the outflow \citep{2014ApJ...793...31S, 2014ApJ...796...29S}

\subsubsection{Diffusivity description of the current paper}
We have investigated a variety of diffusivity prescriptions, all of them can be represented in the following form
for the poloidal magnetic diffusivity,
\begin{equation}
\Ep = \ass (\mu_{\rm D})  \Cs \cdot H \cdot \Fe(z),
\label{eq:diff}
\end{equation}
where the vertical profile of the diffusivity is described by a function
\[ \Fe(z) =   \left\{
\begin{array}{ll}
      1 & z\leq H \\
      \exp(- 2(\frac{z-H}{H})^2 ) & z > H, \\
\end{array} 
\right. \]
that confines the diffusivity to the disk region (see paper I).
The anisotropy parameter $\chi \equiv \Et/\Ep$ quantifies the different strength of poloidal and 
toroidal diffusivity.
It is common to assume $\chi$ of order the of unity. 
For viscous disks \citet{2000A&A...353.1115C} showed that there is a theoretical limit for $\Et$, namely $\Et > \Ep$.
Highly resolved disk simulations indeed suggest $\chi \simeq 2 ... 4$ \citep{2009A&A...504..309L},
implying that the toroidal field component typically diffuses faster than the poloidal component.

Although a parameterization as in Eq.~\ref{eq:diff} is commonly used (except for a different profile function $\Fe(z)$
and the thermal scale height $H_{\rm T}$ instead of $H$ as in our case), 
there are no clear constraints upon the value $\ass$ may take. 
Recent numerical modeling of the MRI applying a non-zero net magnetic field indicate rather high values, 
$\ass \simeq 0.08 - 1.0$, 
with a corresponding magnetization $10^{-4} , 10^{-2}$
(see \citealt{2013ApJ...767...30B} and our discussion in paper I).
Obviously, different functions of $\ass (\mu_{\rm D})$ will lead to a different disk 
evolution.
We start from a general prescription for magnetic diffusivity applied by other authors before 
\citep{2004ApJ...601...90C, 2007A&A...469..811Z, 2012ApJ...757...65S},
\begin{equation}\label{eq:stdiff}
\Ep = \am \Va \cdot H \cdot \Fe(z)
\end{equation}
by applying $\ass = \am \sqrt{2\mu_{\rm D}}$, where $\Va = \Bp / \sqrt{\rho}$ is the Alfv\'en speed,
and $\mu_{\rm D}$, $\Cs$, and $H$ are the magnetization, the adiabatic sound speed and the local disk height,
respectively, measured at the disk midplane. 
We evolve $\ass$ and $\Cs$ in time, but for the sake of simplicity we keep $H$ and
$\Fe(z)$ constant in time, thus equal to the initial distribution.

The majority of simulations in the literature consider a magnetic field strength in equipartition with the
gas pressure.
Studying weakly magnetized disks, we have found that there exists an upper limit for the anisotropy 
parameter, above which the simulations show an irregular behavior \citep{2014ApJ...793...31S}.
In cases where the vertical velocity term can be neglected (e.g. for a very weak magnetic field with 
$\mu_0 \leq 0.02$),
the anisotropy parameter is $\chi < 1/\am^2$, which for our choice of $\am$ is about 0.4. 
By probing the $\chi$ parameter space we found that in order to obtain a stable accretion-outflow configuration
for weakly magnetized disks, the $\chi$ should be in the range $0.3-0.7$. 
We therefore decided to apply $\chi = 0.5$ for all of our simulations.

\subsubsection{Comparison to literature works}
The diffusivity profile we apply is constant in time and space (while its strength changes in time).
We note that other authors have applied time-dependent diffusivity profiles so far,
in particular adapting the strength and scale height of diffusivity following the
time evolution of the isothermal sound speed and the Alfv\'en speed at the disk
mid-plane.
Such a consideration implies a strong feedback loop for the simulation that may give
rise to an instable disk evaluation. 
This is why we have applied the simplified approach of a constant-in-time magnetic 
diffusivity profile.

In \citet{2010A&A...512A..82M} the magnetic diffusivity is time-dependent and follows the
turbulent viscosity that is updated considering the isothermal sound speed along the disk midplane
(Prandtl number 2/3).
The disk magnetization is not considered for the strength of the diffusivity.
The vertical profile of the diffusivity is not discussed, and also not the numerical procedure how to 
practically update the scale height of the diffusivity distribution\footnote{
It would be very interesting for the understanding of their simulations to see the time evolution 
of the diffusivity distribution (numerical values or maps).
Further, the authors do not elaborate on their application of magnetic diffusivity on the scaled 
outer grid in their simulations.
Because the user manual for PLUTO explicitly states that magnetic diffusivity should {\em not} be applied on
stretched grids, the method implemented in \citet{2010A&A...512A..82M} may be of use to the community at large.
}

\citet{2009MNRAS.400..820T} apply a magnetic diffusivity depending on the thermal
scale height $H$ and the Alfv\'en speed $V_{\rm A}$, both calculated for the disk-midplane,
$\eta_{\rm m} \propto H V_{\rm A} \exp(-2 z^2/H^2)$.
The authors do not further specify whether and how they update the magnetic diffusivity
in time, the diffusivity distribution is described only as an initial condition.

In a previous publication \citep{2012ApJ...757...65S, 2013ApJ...774...12F} we have in particular 
investigated how the diffusivity profile affects jet launching and in particular mass loading.
In \citet{2013ApJ...774...12F} we have suggested a new time-dependent model for magnetic diffusivity
that depends from the {\em local} disk pressure and magnetization.
In particular the time evolution of the magnetic diffusivity has been shown. 

On the other hand, the application of a diffusivity profile constant in time may have its caveats.
That is that the disk may vertically shrink (quantified by the thermal disk height $H_{\rm T}(r,t)$),
while the diffusivity remains vertically fixed with a fixed scale height for the diffusivity
profile $H$.
This may have a strong effect on the mass loading and, thus, on the jet dynamics.

A strong disk magnetic field may compress the disk hydrodynamic structure, however, this also depends
on the field curvature and not only on the field strength.

Direct simulations of disk turbulence have shown that the disk turbulent diffusivity profile may
indeed extend beyond the thermal scale height \citep{2010MNRAS.405...41G}.

In summary, we believe that our choice of diffusivity indeed provides physically meaningful results.
We see as an advantage of our diffusivity profile that we may control its functional form during
the simulation at any time.
We thus know its spatial distribution and may disentangle physical effects that depend on the local
value for the diffusivity.
Our diffusivity profile allow for a more stable simulation, as it allows for mass accretion
and jet mass loading also when the disk thermal scale height becomes low.

From the simulations in the literature cited above, it becomes not clear, how the diffusivity
profile actually looks like during the time evolution.
In particular, when the thermal disk height becomes small, numerical diffusivity may affect the
physical diffusivity such that the initially chosen functional form of diffusivity may actually 
not hold.
We also believe that the grid resolution is usually not sufficient to allow for a physical
treatment of the feedback between the disk midplane Alfv\'en velocity and the vertical profile
of the disk turbulence that governs the magnetic diffusivity.

\begin{figure}
\centering
\includegraphics[width=9cm]{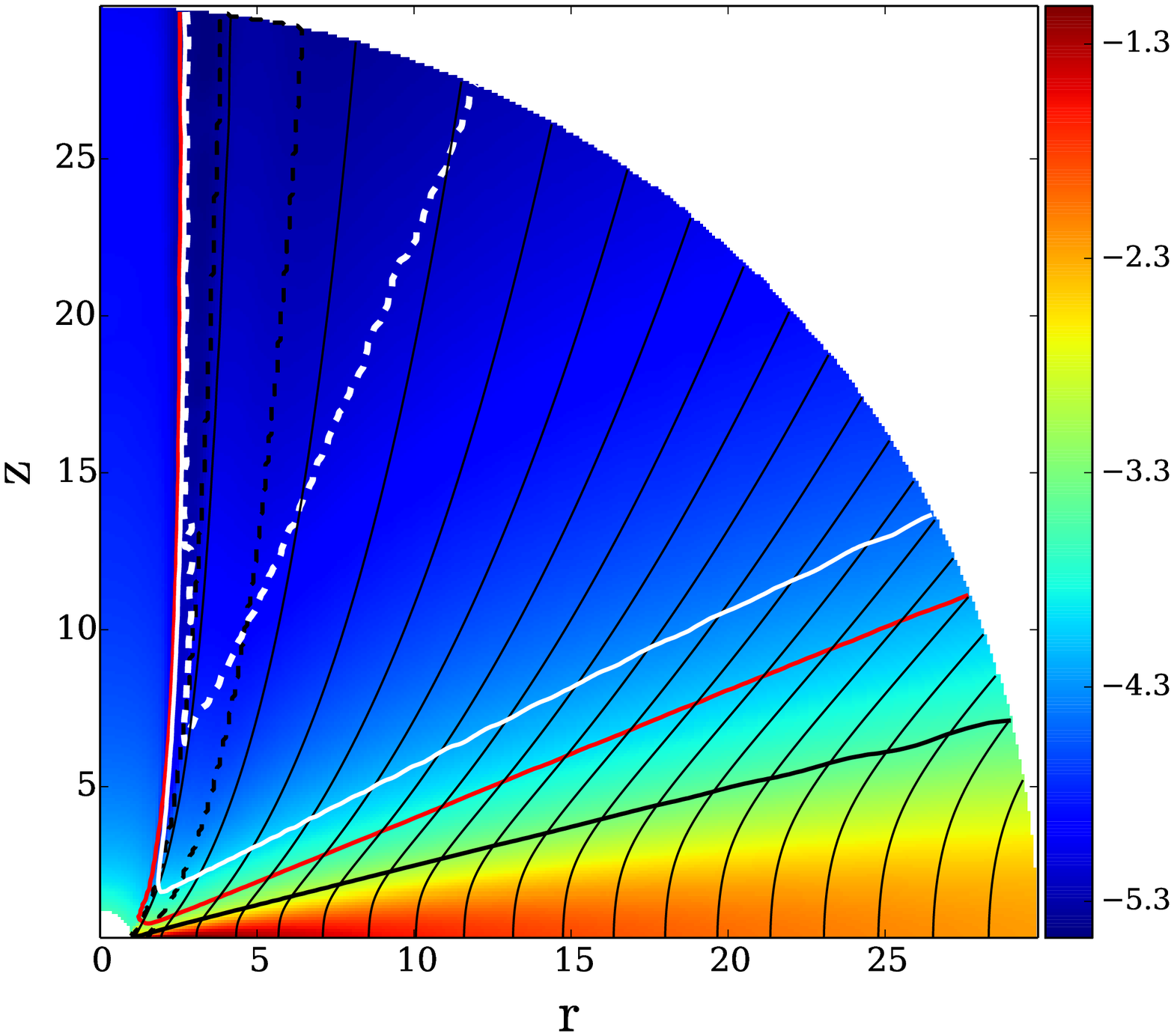}
\includegraphics[width=9cm]{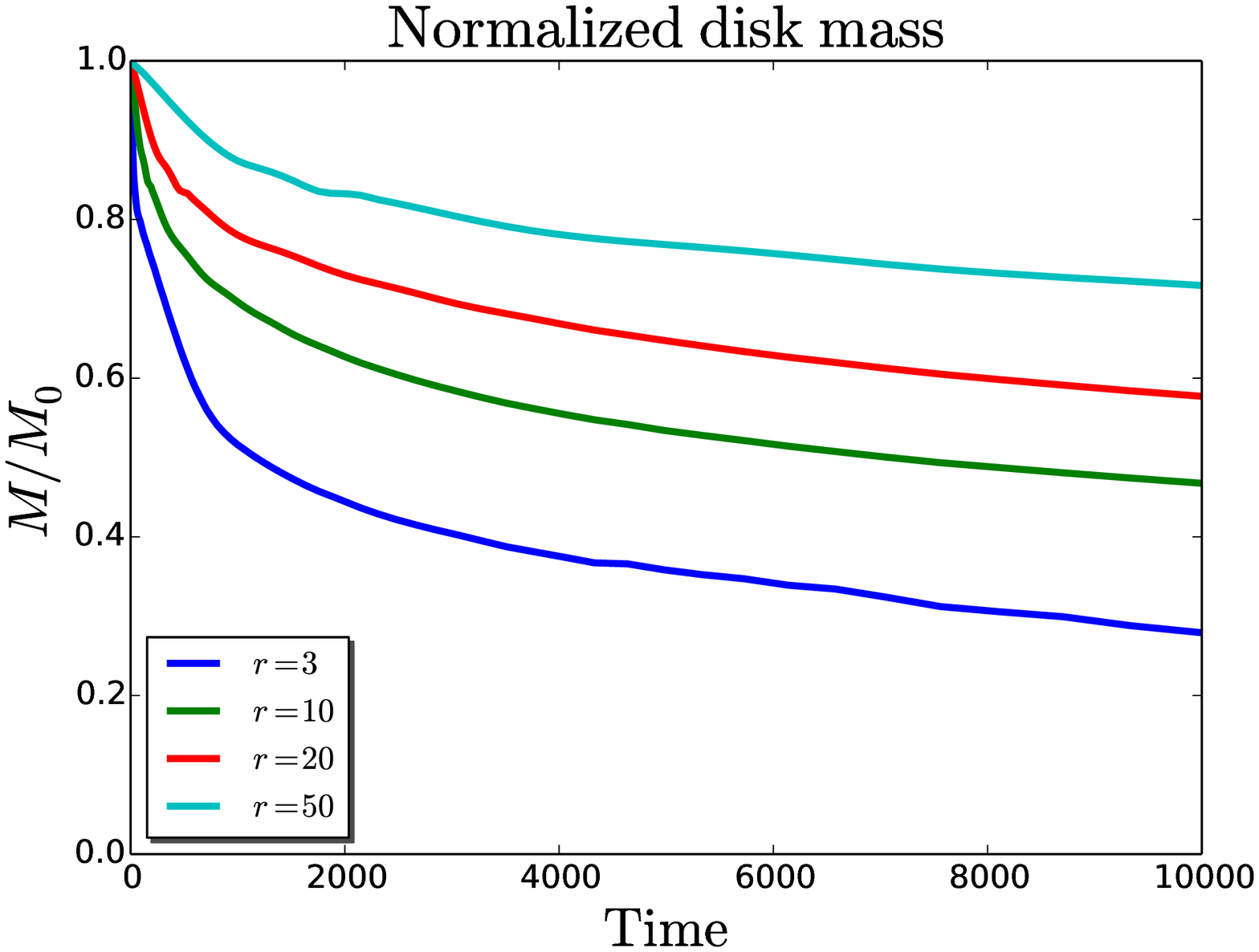}
\caption{Typical structure of the disk-jet system at $T=10,000$ (top).
Colors show logarithmic density. Thin black lines show the magnetic field lines. 
Thick dashed lines, rooted at $r = 1.1$ and $r = 1.5$, mark the jet region, investigated
further in the paper. 
Lines mark the disk surface (black), the sonic surface (red) the Alfv\'en surface (white),
and the fast surface (dashed white). The jet sheet further investigated below is indicated
by dashed black lines.
The time evolution of the disk mass included within increasing radius decreases with time (below).
}
\label{fig:jet_struct}
\end{figure}

\subsection{Simulation parameters}
The magnetic diffusivity $\eta = \am f(R,\Theta)$ consists of the magnetic diffusivity
parameter $\am$ that defines its strength and a function for the profile $f(R,\Theta)$ (see above). 
The anisotropy parameter is chosen $\chi = 0.5$.
In general the magnetic diffusivity also defines the trend for the disk magnetization - magnetic field 
amplification by advection versus magnetic field decline by 
diffusion\footnote{This balance can be denoted by the magnetic Reynolds number
$Re_{\rm M}\equiv V L/\eta$ that relates dynamical and diffusive effects by comparing typical 
velocities $V$ and length scales $L$ to the diffusivity $\eta$.}.
The resulting magnetization - measured as a snapshot in time - determines the 
actual launching conditions.

The magnetic diffusivity parameter $\am$ turns out to be crucial for obtaining a 
quasi-steady state solution.
As discussed in paper I, there is a critical magnetic diffusivity parameter $\am$ such 
that for our standard diffusivity prescription the diffusive and advective processes are in balance.

For our setup, we find that $\am \approx 1.6$ is a good choice for a smooth and 
long-term evolution of the simulation.
However, a slight deviation from the critical value for the diffusivity parameter leads the simulation into
a state that is dominated by either diffusion or advection.
As the simulation progresses, this deviation amplifies (see paper I) and the actual quantities of the disk change.
Essentially, this process allows us to investigate the disk-jet evolution over a wide range of the actual disk
magnetization $\mu =  2\times 10^{-4} - 3\times10^{-0.5} $, 
since the magnetic flux of the disk can be either advected inwards increasing the disk
magnetization, or it can diffuse out leading to a lower disk magnetization.

When the actual disk magnetization is in the above mentioned range, 
all simulations show a robust behavior over long time scales and can be smoothly evolved for more 
than 10,000 time units (corresponding to 1600 disk revolutions at the inner disk radius),
until we stop the simulation. 
We are able, however, to perform simulations for more than 500,000 time steps (see paper I).

Table~\ref{tbl:simuls} summarizes all simulations discussed in this paper.
We denote our simulations in alphabetical order according to the strength of the 
{\em initial} magnetization.
The plus and minus signs indicate a slightly higher or lower diffusivity $\am$ with respect 
to each other.
We note, that a {\em lower} diffusivity parameter leads to {\em stronger} actual disk magnetization,
as a result of faster advection and accumulation of the magnetic flux.

\begin{table*}
\caption{Parameter runs of our simulations, alphabetically labeled according to the initial disk 
magnetization at the inner disk radius $\MUzero$. 
The plus and minus sign indicate a slightly higher or lower diffusivity, respectively, 
denoted by the parameter $\am$ (see Equation 10).
}
\begin{center}
    
\begin{tabular}{llllllllllll}
   ~    &     A &    B+ &     B &    C+ &    C- &    D+ &    D- &   E+ &   E- &    G &   H \\
\noalign{\smallskip}    
\hline \hline    
\noalign{\smallskip}
$\mu_0$ & 0.001 & 0.002 & 0.002 & 0.003 & 0.003 & 0.005 & 0.005 & 0.01 & 0.01 & 0.02 & 0.03 \\
$\am$   & 1.6   & 1.6   & 1.5   & 1.65  & 1.6   & 1.6   & 1.55  & 1.65 & 1.6  & 1.8  & 1.65 
\end{tabular}
\end{center}
\label{tbl:simuls}
\end{table*}

\subsection{General jet-disk evolution}
In paper I, we have presented an approach that allows us to study a broad parameter space 
of the disk-jet structure.
While in paper I our discussion concentrated more on the disk evolution,
in the present paper we are interested in the evolution of the outflow dynamics and 
how it correlates to the physical disk properties.
In the following we briefly comment on the general evolution of the disk-outflow dynamics, mainly 
summarizing our simulations of paper I.

After starting the simulation from the initial conditions, the disk evolves into a new
equilibrium.
When the disk wind establishes, the efficient removal of angular momentum from the disk by the
wind leads to accretion of disk material.
A new dynamical state of the disk will be reached - defined by the balance
between inward advection of mass and magnetic flux and outward diffusion of the magnetic flux.

The disk material is lifted vertically and couples to the magnetic field lines of jet and outflow.

The evolution of the accretion-ejection structure finally reaches a quasi steady state
that is further affected by the mass loss due to accretion and ejection. 
Thus, the disk structure evolves slowly in time, while the outflow that is launched
during quasi steady state, establishes on a much shorter time scale - that is the 
Keplerian time scale at the foot point of the outflow.

Figure~\ref{fig:jet_struct} shows a small subset $(z<30$) of a large numerical grid of a typical 
simulation ($z<1500$, see paper I).
Here the structure of the disk-jet system has evolved until time $T = 10,000$.
The location of typical surfaces, namely the disk surface, defined as the surface where the radial 
velocity changes sign, $\Vr=0$, the sonic surface and the fast magnetosonic surface,
have become stationary, indicating that the disk-outflow structure in this volume has reached a
quasi-steady state.
We denote this evolutionary stage as a {\em quasi-steady} state, since there is still a very slow evolution 
of the system is for long time scales.
This is because the inner part of the disk experiences a slow mass depletion by the accretion and ejection.
Furthermore, it takes much longer for the outer part of the disk to evolve and thus to reach a steady state.

For the setup with the diffusivity parameter $\am$ chosen about its critical value, 
it takes about 1000 dynamical time steps (thus 160 inner disk revolutions) 
to reach a quasi-steady state.
The quasi-steady state situation is determined by the balance between accretion and ejection 
and also by the mass transfer rate from the  disk to the jet.

The magnetic diffusivity acts to level out the magnetic field gradient, thus setting the 
overall structure of the magnetic field.
However, one should keep in mind that the magnetic field strength and structure of real accretion 
disks are also not known.
In our approach, we will compare the disk magnetization and the outflow characteristics
that are both influenced by the magnetic diffusivity.
As we will see, another diffusivity prescription will probably reach a different stage of 
disk magnetization at a different time, however, the jet characteristics we find from the
simulation, are just dependent on the actual magnetization at the jet foot point.
They are thus memoryless and do not depend on the {``}history{''} of how this state has been reached.

In order to correlate the jet dynamics to the underlying disk, we will apply 
the steady-state MHD conservation laws along the field lines.
In general, these MHD integrals are conserved only in a steady-state, axisymmetric MHD.
In our simulations we always reach a quasi-steady state for the high-speed jet component and
the inner disk structure from where this jet is launched.
In the case of a low disk magnetization ($\MUdisk \le 0.001$), we find that the jet becomes 
slightly disturbed above the Alfv\'en surface.
We believe that these perturbations are presumably triggered by the shear between 
the axial flow injected from the coronal boundary (as a boundary condition) 
and the jet that is physically launched from the inner disk.
However, since the outflow remains in a quasi-steady state until it reaches the Alfv\'en surface, 
we may indeed consider the MHD integrals obtained at this location and interrelate 
them to the underlying disk magnetization.
Therefore the correlations we obtain are valid also for the low magnetization case.
In the case of a high disk magnetization, the outflow remains unperturbed and the MHD integrals 
remain conserved along the jet also for larger distances from the launching point.

Figure \ref{fig:jet_struct} (bottom) shows also the temporal evolution of the disk mass.
We see that after a sharp initial decline, the disk mass is still deceasing slowly due to the mass
loss by accretion into the inner boundary and ejection into the jet.
For later evolutionary times the disk looses about 10\% of its mass over 6000 dynamical time steps
for e.g. $r<10$,
so the typical time scale when half of the disk mass is gone would be $\tau_{\rm disk} \simeq 30,000$
dynamical time steps.
This time scale would be the time scale on which the disk evolution happens, as seen by the
evolution of the disk magnetization\footnote{This is why we had to develop another magnetic diffusivity 
model for the very long-term evolution simulations (more then 500,000 dynamical time steps) in paper I, 
allowing for a sufficiently rapid mass replenishment from the outer disk to the inner disk.}.

The disk diffusive time scale is much faster. 
For a typical magnetic diffusivity $\eta \simeq 0.03$ the time scale for magnetic diffusion is 
$\tau_{\eta} \equiv l^2 / \eta \simeq 3,000$ if we consider the jet launching area $l\simeq 10$.
Previous simulations have show that a quasi steady-state between advection and diffusion of magnetic 
flux is reached (see discussion above and \citealt{2012ApJ...757...65S}.
This time scale is similar to the advection time scale 
$\tau_{\rm adv} \equiv l / v_{\rm acc} \simeq 1,000$, applying typical accretion velocities we
measure in our simulation of $v_{\rm acc} \simeq 0.01$.

The kinematic time scale for jet formation usually scales with the Keplerian time scale at the jet 
foot point - simply due to the fact the the jet typically reaches asymptotically the same velocity
$\tau_{\rm jet} \simeq l_{\rm grid} / v_{\rm jet} = l_{\rm grid} / v_{\rm Kep}$. 
Typically, $\tau_{\rm jet} \simeq 160$ for a grid size of 1000 inner disk radii and a jet launched
from close to the inner disk radius.
The overall structure of the outflow - in particular its collimation - depends, however, on the 
{\em overall} pressure and magnetic field distribution, and, thus, also on those parts of the 
outflow that are launched from the outer disk and that evolve on much larger Keplerian time scales.
We may define a time scale considering the whole outflow structure 
$ \tau_{\rm outflow} \simeq  l_{\rm grid} / v_{\rm Kep,grid}$.
For the parameters used above we find  $ \tau_{\rm outflow} > 30,000\,\tau_{\rm jet}$.

\begin{table}
\caption{Comparison of typical times scales in the simulation runs.
The exact values depend on the radius and may also somewhat change in time.
Time unit is $T_0 = R_0 / V_{\rm K,0}$. In code units $T_0 = 1.0$.
Shown are the jet kinematic or propagation time scale $\tau_{\rm jet}$, 
the diffusive time scale $\tau_{\eta}$,
the advection time scale $\tau_{\rm adv}$, both considering a scale length of $l=10$ for the jet launching
area, the disk {``}life time{''} $\tau_{\rm disk}$ during which the disk looses about 50\% of its mass,
and the time scale for the whole disk outflow to establish a dynamical steady state, $\tau_{\rm outflow}$,
that is the Keplerian time scale for the outer disk radii. 
}
\begin{center}
\begin{tabular}{llllll}
time scale & $\tau_{\rm jet}$ &  $\tau_{\eta}$ & $\tau_{\rm adv}$ & $\tau_{\rm disk}$ & $\tau_{\rm outflow}$ \\
    \noalign{\smallskip}    \hline \hline    \noalign{\smallskip}
 time in $t_{\rm in}$ & 160 &  3,000 & 1,000 & 30,000 &  $5\times10^6$ \\
\end{tabular}
\end{center}
\label{tbl:timescales}
\end{table}

\section{Characteristic outflow properties}
In paper I, we have explored the general structure and evolution of the simulations
with emphasis mainly on the disk variables, namely accretion-ejection efficiency, 
and the corresponding fluxes of mass and energy.
In this paper, our main concerns are the jet quantities, especially the jet integrals
that can be compared to the steady-state MHD theory,
and the strong correlation between the jet physical
properties and the disk magnetization.

\begin{figure}
\centering
\includegraphics[width=9cm]{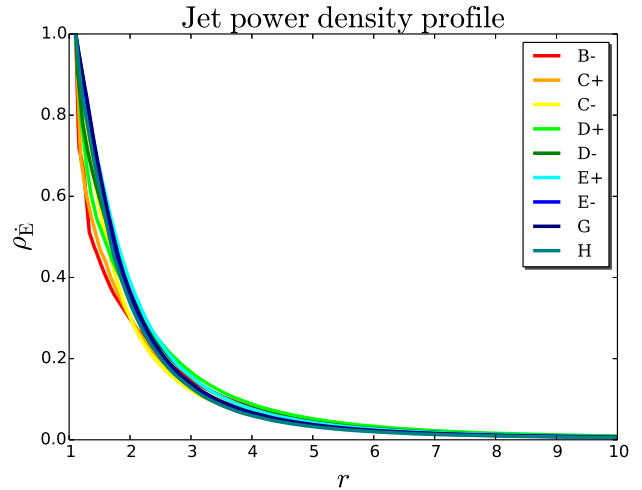}
\caption{Profile of the jet power density $\rho_{\rm \dot{e}}$ at $T=10,000$ 
along the surface $z = H(r) + 1$ (parallel to the disk surface), 
normalized to the value at the inner edge of the surface at $r=1.1$, $z=1.22$, for a range of parameter runs 
(see Table~\ref{tbl:simuls} for the notation).
}
\label{fig:jet_rpow}
\end{figure}

\subsection{Steady-state jet MHD integrals}
\label{sec:integrals}
Below we compare the dynamical parameters of the outflow in a quasi-steady 
state with the parameters of the classical steady-state MHD wind theory.
In axisymmetric, stationary, ideal MHD there are four quantities that are conserved 
along the magnetic flux $\Psi \equiv \int \vec{B}_{\rm p} \cdot d\vec{A}$, 
namely the mass loading per flux surface enclosing the magnetic flux $\Psi$,
\begin{equation}\label{eq:kint}
k (\Psi) =  \frac{\rho \Vp}{\Bp},
\end{equation}
the angular velocity of the field lines
\begin{equation}\label{eq:wint}
\om (\Psi) =  \frac{\Vphi}{r} - \frac{k \Bphi}{\rho r}, 
\end{equation}
the specific angular momentum
\begin{equation}\label{eq:angmom}
l (\Psi) = l_{\rm V} + l_{\rm B} = r\Vphi - \frac{r \Bphi}{k}
\end{equation}
(with the kinematic and magnetic contributions  $l_{\rm V}$ and $l_{\rm B}$, respectively),
and the specific energy
\begin{equation}\label{eq:energy}
e (\Psi) = \frac{\Vphi^2 + \Vp^2}{2} + \Grav + h + (l - r \Vphi) \om.
\end{equation}
There are two more derived MHD integrals that are commonly used, namely
\begin{equation}\label{eq:jint}
j(\Psi) = e - \om l,
\end{equation}
that contains only kinematic quantities, and the maximum jet speed
\begin{equation}\label{eq:vinf}
V_{\rm inf} = \sqrt{2 e},
\end{equation}
that is the asymptotic speed of a propagating jet.
For further details and derivations concerning the steady-state MHD jet integrals we refer 
to the review paper of \citet{2007prpl.conf..277P}.

In order to explore the correlation of these integrals (calculated along the outflow)
with the physical quantities describing the disk (mainly the actual disk magnetization), 
we normalize the integrals to their midplane values.
In this paper, all jet integrals are presented in non-dimensional units, namely 
\begin{equation}\label{eq:allint}
\tilde{k} \equiv \frac{k}{\sqrt{\rhod}}, \,
\tilde{w} \equiv \frac{\om}{\Omega_{\rm F_{\rm D}}},\,
\tilde{l} \equiv \frac{l}{l_{\rm D}}, \,
\tilde{e} \equiv \frac{e}{e_{\rm D}}, \,
v_{\rm inf} \equiv \sqrt{e}, \,
\tilde{j} \equiv \frac{j}{j_{\rm D}},
\end{equation}
where the index {``}D{''} denotes the value calculated at the disk midplane (at a certain radius).
In the following we will omit the tilde sign for simplicity\footnote{Note that we apply the
notation \citet{2007prpl.conf..277P} and \citep{1992ApJ...394..117P}. 
This differs somewhat from the notation used e.g. in \citet{1982MNRAS.199..883B} or \citet{2000A&A...353.1115C},
who denote the mass loading per flux surface with $\kappa$ and the specific angular momentum with $\lambda$.}.
Note that the underlying disk remains (sub-) Keplerian at the midplane during all simulations, 
thus $\Vphi \propto r^{-0.5}$.
Therefore, the $\omz, l_{\rm D}, e_{\rm D}, j_{\rm D}$ are constants in time, 
but the density $\rhod$ varies as disk evolves.
For example, $l=3$ corresponds to the jet layer whose {\em specific} angular momentum 
is three times higher than the specific angular momentum of the underlying Keplerian disk 
at the jet layer foot point (at the midplane).

\subsection{How to quantify the disk and jet properties}\label{sec:djquan}
The jet integrals are essential outflow properties that potentially connect the asymptotic 
outflow to the disk properties.
Naturally, the numerical results are calculated on numerical grid points.
Thus, some averaging procedure is required to derive the jet integrals 
along a magnetic flux surface that is intersecting the grid cells.
Here, we briefly describe how we calculate the integrals from the numerical
simulation. 
Such a prescription has not always been provided in the literature.

We obtain the number value for each jet integral by averaging across an outflow sheet of 
certain width and length located just beyond at the  Alfv\'en surface.
This average is performed as a snapshot for a certain time.
For each of the integrals we have explicitly checked whether it is conserved along the 
corresponding sheet along the outflow.

The extent of the averaging area should be at one hand narrow enough in order to provide
a meaningful value for the flux surface.
On the other hand it must be large enough to provide a accurate average value.
We consider a jet sheet that is confined between two {``}adjacent{''} magnetic flux surfaces
and that is rooted in the innermost disk between $r = 1.1$ and $r = 1.5$.
From here the most powerful part of the jet is launched.
The radius $r = 1.1$ is well separated from the inner boundary by eight grid cells.

For the length of the averaging domain, we take a {"}box{"} ranging from the
Alfv\'en point of that flux surface $z_{\rm A}$ to $(z_{\rm A}+2.0)$.
Depending on the vertical position along the flux surfaces, the radial separation 
between the flux surfaces varies (it is $\Delta r=0.4$ only at the equatorial plane).

We will refer to $X_{\rm D} = X(0)$ as the disk quantities (at the foot point of the jet), 
while $X_{\rm A} = X(z_{\rm A})$ denote the jet quantities.
By computing average values over a bundle of magnetic flux surfaces 
(a jet sheet) we are able to provide a robust measure for the integrals.

This can be demonstrated by e.g. exploring the radial profile of different physical quantities 
along the disk surface - thus by comparing the properties of different sheets of the jet.
As an example, Figure~\ref{fig:jet_rpow} shows the radial profiles of the jet power density 
$\rho_{\rm \dot{e}} =  \rho e \Vp$ along a surface $z = H(r) + 1$ parallel 
to the disk surface\footnote{Note that all different the lines correspond to different simulation runs}.
These profiles were all plotted at the same time, $T=10,000$. 

We find that all profiles follow a very similar shape, thereby approving our average measure for 
the jet parameters along the inner most jet sheet as indeed representative for the jet as a whole.
Note that the underlying disk properties are completely different as resulting from to different 
simulations which started from different initial conditions and evolve differently.
In particular, the actual magnetization, the density distribution or the accretion velocity 
are different for each of the simulations.

\subsection{Conserved quantities along the jet}
In this section, we discuss how the physical jet properties vary along the outflow.
We follow the outflow from its foot point to the asymptotic domain.
Note that when referring to the z-axis in following figures, all quantities are averaged over 
the jet layer, although the plot is along the z-axis.

\begin{figure}
\centering
\includegraphics[width=9cm]{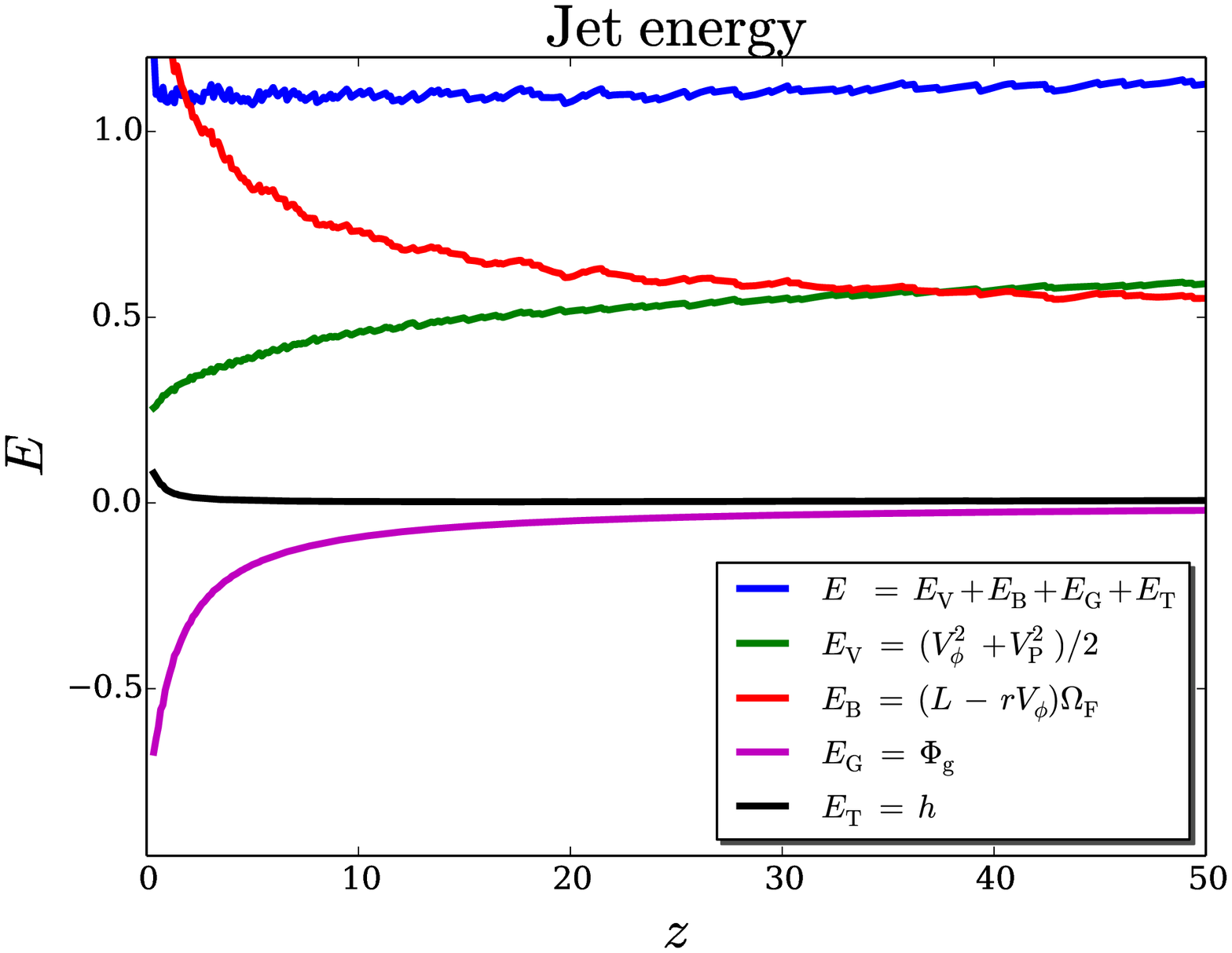}
\includegraphics[width=9cm]{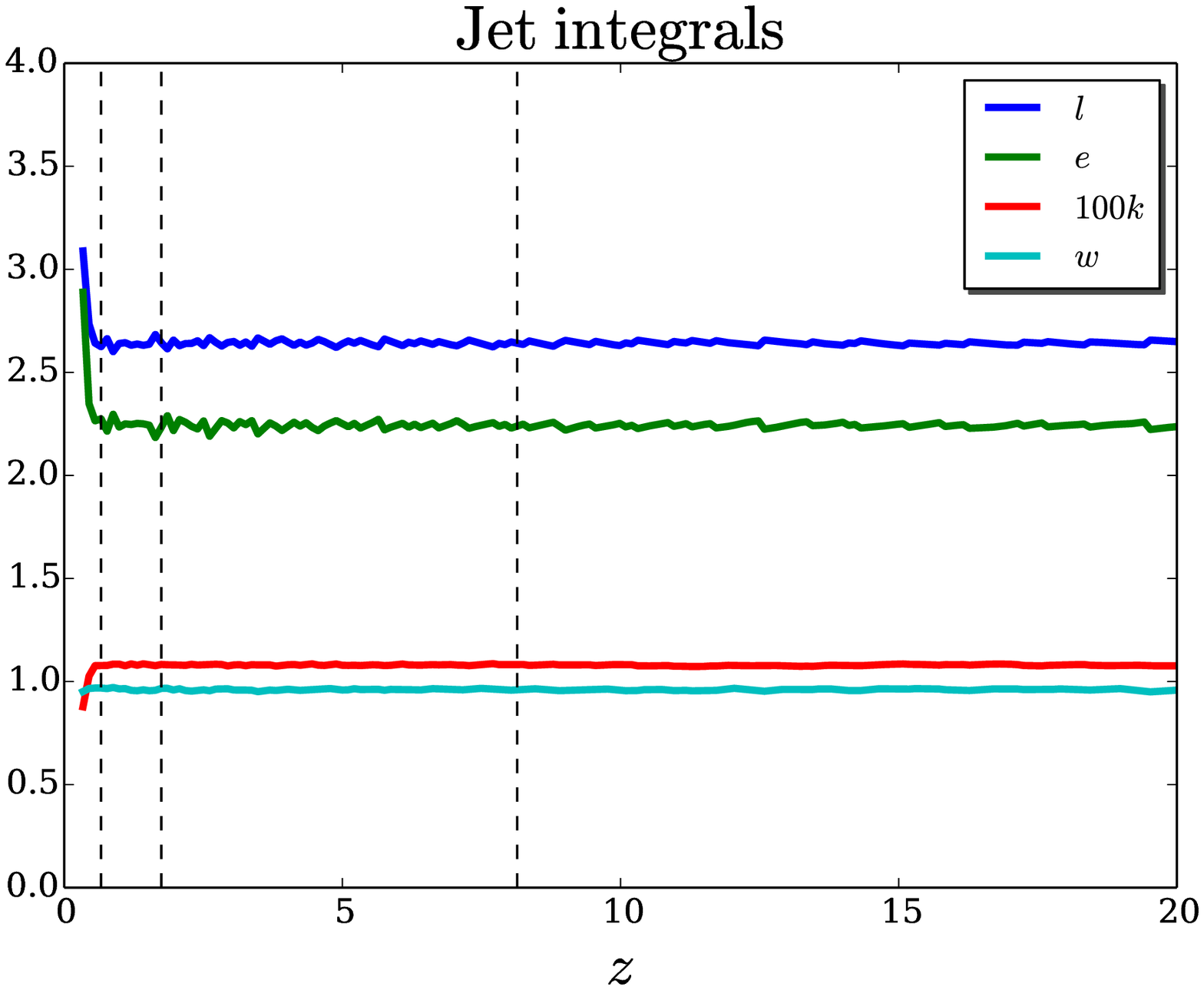}
\caption{Jet specific energy contributions along the vertical direction $z$ for the
simulation {\em B} simulation at time $T=10,000$ (top).
Different energy components are indicated by colors:
total (blue), kinetic (green), magnetic (red), gravitational (magenta), and thermal (black) energy.
Jet integrals along the magnetic field line rooted in the innermost disk are
for simulation {\em E+} time $T=10,000$ (bottom).
Shown are the jet specific angular momentum $l$, the mass load (multiplied by a factor 100), the field line angular 
velocity $\omega$, and the jet specific energy $e$ (normalized as described in Eq.~17).
Vertical lines indicate the position of the sonic surface / slow magnetosonic surface
(indistinguishable), the Alfv\'en surface, and the fast magnetosonic surface, respectively.
}
\label{fig:jet_ze}
\end{figure}

Figure~\ref{fig:jet_ze} (top) shows the profiles along the jet for the different 
jet specific energy components for simulation run {\em B}.
We see that although we have averaged the variables 
within a jet layer containing a bunch of magnetic field lines, the total energy 
$e = e(\Psi(r,z))$ is well conserved along the jet, approving our averaging approach.
The same is true for all jet integrals studied in this paper.
The thermal energy is negligible everywhere in the jet.
The gravitational energy is negligible only far from the disk ($Z > 20$).
Here the outflow velocity has surpassed the local escape speed 
 $\Vp(R,\theta) \gg V_{\rm escape}(R,\theta)$
and has become gravitationally unbound to the central object.
Figure~\ref{fig:jet_ze} (top) further indicates how the magnetic energy of the jet is
transferred into the jet kinetic energy.
Far from the disk the transformation of the magnetic energy into the kinetic energy 
becomes less efficient. 
This decrease of acceleration is related to the increase of collimation.
  
Figure~\ref{fig:jet_ze} (bottom) shows the profiles along the jet for the 
the other MHD integrals.
Shown are the jet angular momentum $l$, the mass load (multiplied by a factor 100), 
the field line angular velocity $\omega$, and the jet energy $e$.
Vertical lines indicate the position of the magnetosonic surfaces.
The sonic surface and the slow magnetosonic surface are indistinguishable.
the Alfv\'en surface, and the fast magnetosonic surface, respectively.
The bottom figure demonstrates that our approach for averaging discussed above
is working fine, as all integrals follow a straight line already at the Alfv\'en 
surface.

\begin{figure}
\centering
\includegraphics[width=9cm]{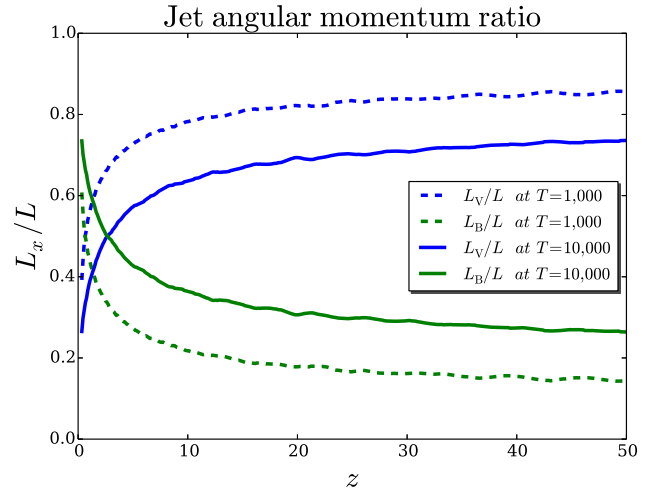}
\caption{Jet angular momentum components along the jet, obtained for the  {\em B} 
simulation.
Kinetic (blue lines) and magnetic (green lines) to total jet angular momentum ratios are 
plotted at $T=1000$ (dashed lines) and $T=10,000$ (solid lines).}
\label{fig:jet_zlx}
\end{figure}

Figure~\ref{fig:jet_zlx} indicates how efficient the magnetic angular momentum 
is transferred into kinetic angular momentum.
We show the angular momentum balance along the jet for two different evolutionary times.
We see that the rate of the angular momentum transfer along the jet is different for 
different evolutionary times. 
At earlier times ($T=1,000$) the magnetic angular momentum is transferred faster 
into the kinetic angular momentum compared to later times ($T=10,000$).
However, what is actually happening is that between these time steps the underlying disk magnetization has 
substantially changed from 
$\MUdisk \approx 0.02$ at $T=1,000$ to $\MUdisk \approx 0.2$ at $T=10,000$.

From our simulations we may therefore conclude that for jets launched from a 
stronger magnetized disk it takes more time to convert all magnetic energy into 
jet kinetic energy.
Further investigations are needed to clarify this point.

\section{The disk - jet connection}\label{sec:dj}
In this section, we obtain a number of correlations between the jet
physical properties and the properties of the accretion disk.
Our main motivation is the question {\em what kind of disks drive what kind of jets?}.

As discussed in Section~\ref{sec:djquan}, the properties of the accretion disk can be specified by 
a few parameters that are quantified at the midplane of the inner disk.
On the other hand, the parameters of the jet can be specified by the MHD integrals that are calculated by
averaging across the inner jet layer at the location of the Alfv\'en surface (thus in the 
ideal MHD region of the outflow).

The material along the midplane is accreted through the inner boundary and is not ejected into the jet.
However, the disk-jet system is tightly connected and the dynamics of both components follow a 
common global pattern.
Even though the inner disk midplane and the outflow at the Alfv\'en point are not directly connected by the 
exchange of material itself, they are in causal connection due to magnetohydrodynamic waves and forces.
The evolution of disk and outflow is interrelated and cannot be separately treated.
It is therefore not surprising that all jet integrals discussed in the Section~\ref{sec:integrals} 
can be, as we will see later, obtained by knowing the disk magnetization.

\begin{figure*}
\centering
\includegraphics[width=8cm]{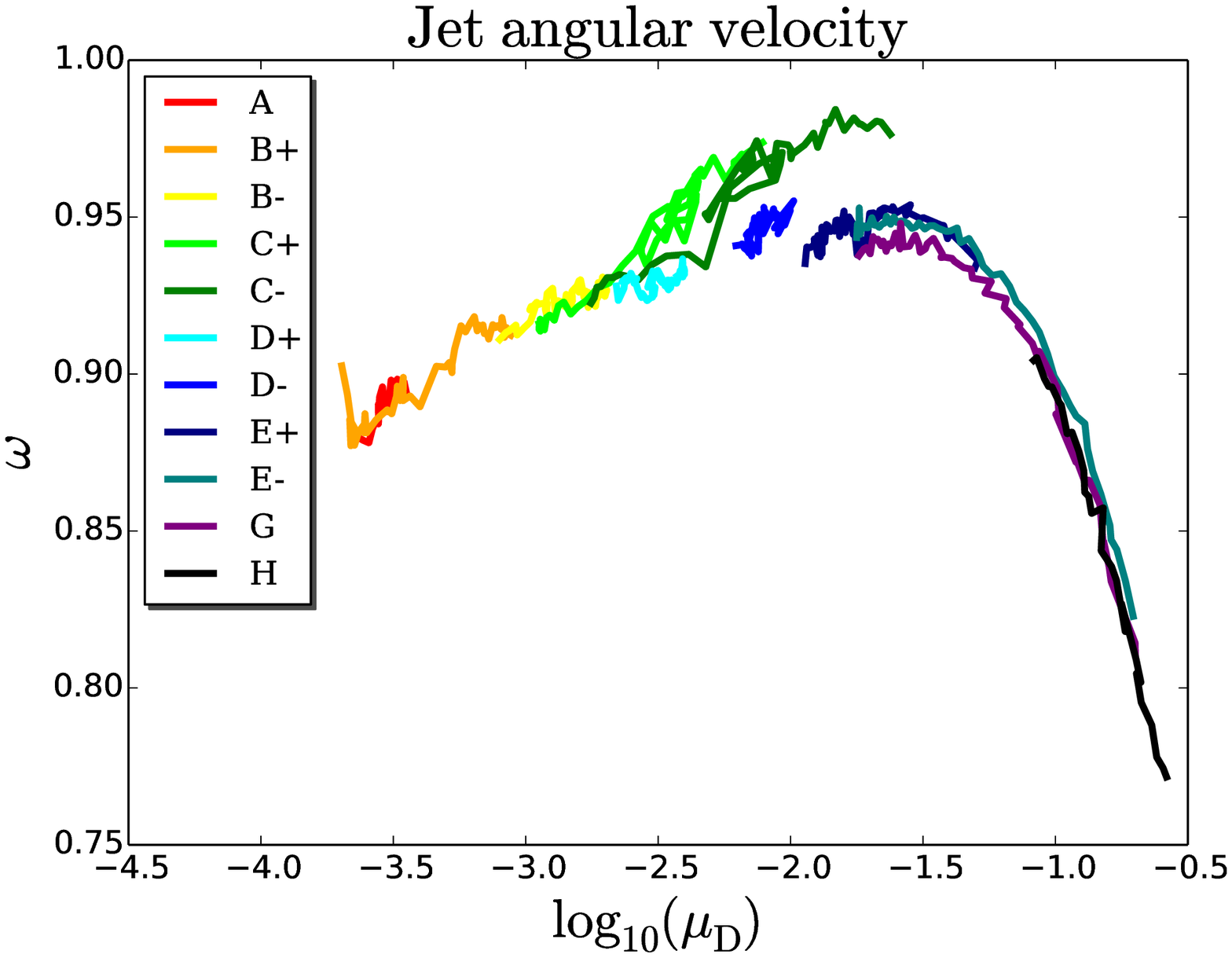}
\includegraphics[width=8cm]{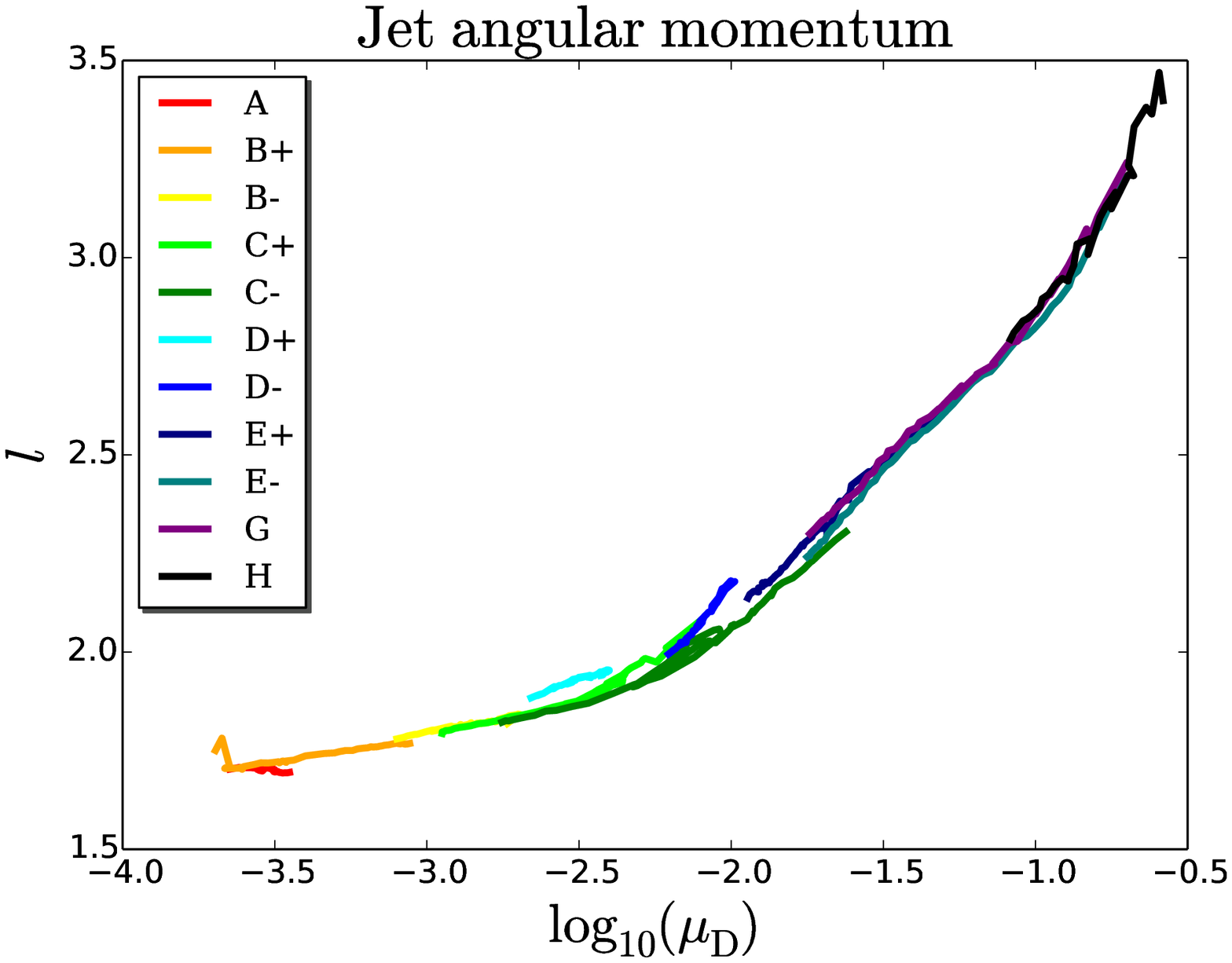}
\includegraphics[width=8cm]{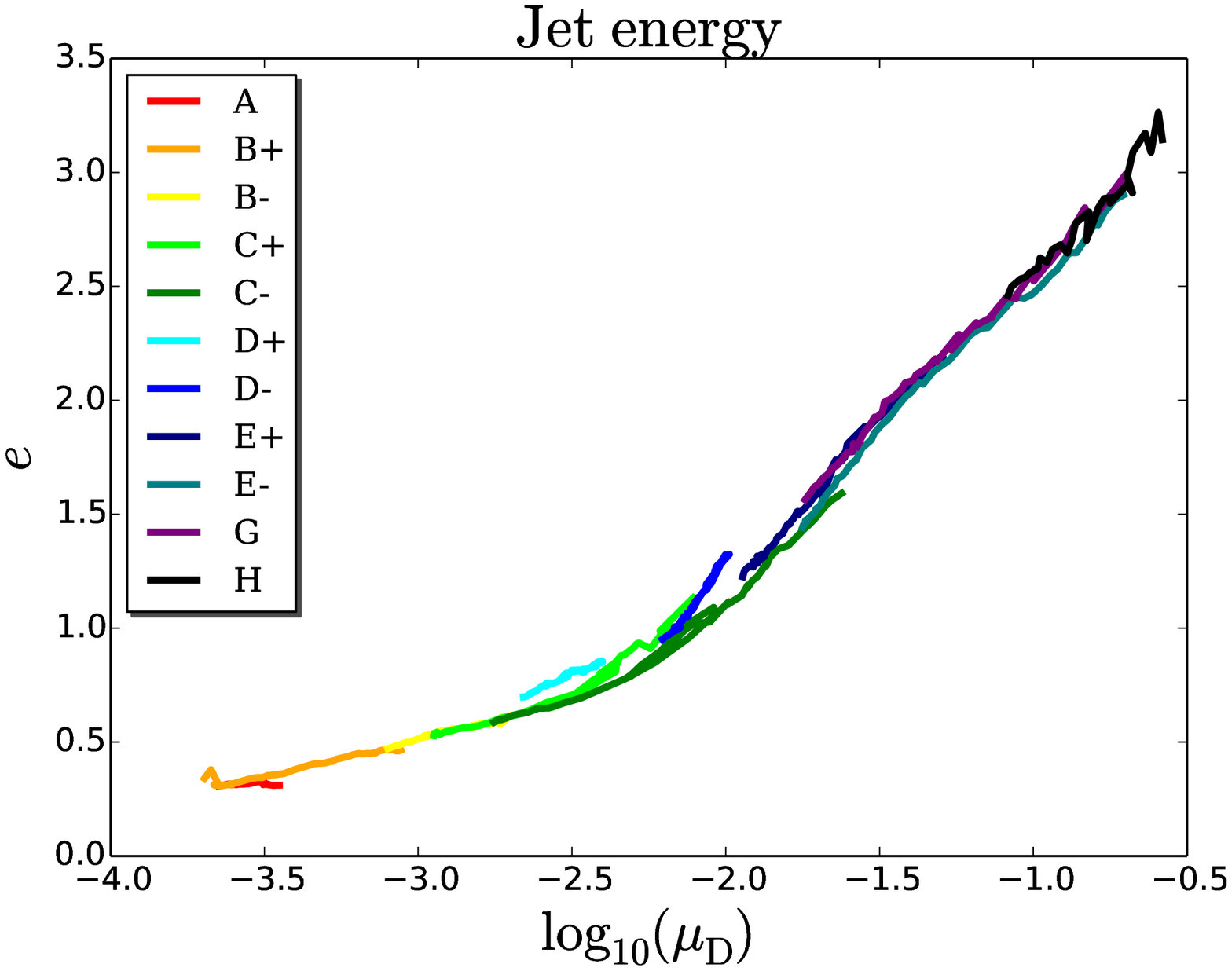}
\includegraphics[width=8cm]{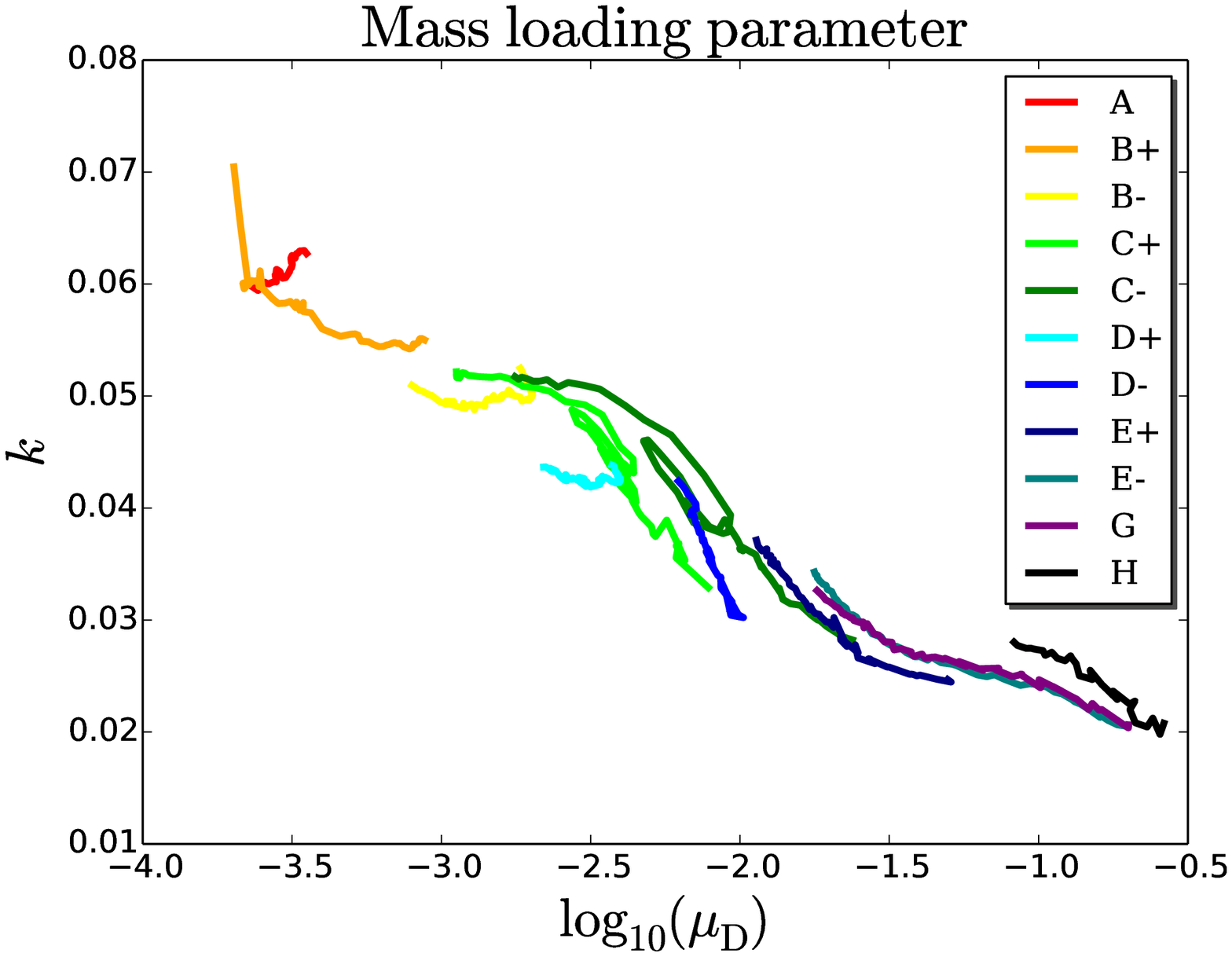}
\caption{Jet physical properties with respect to the disk magnetization $\MUdisk$.
Shown is the jet angular velocity $\omega$,
the specific angular momentum $l$,
the jet energy $e$, and 
mass loading parameter $k$.
Each line represents the evolution of a single simulation (see Table~\ref{tbl:simuls})
from 700 to 10,000 time units.}
\label{fig:jet_mud}
\end{figure*}

Figure~\ref{fig:jet_mud} shows the correlations between the disk magnetization and the 
jet angular velocity, the total (specific) jet angular momentum, the specific jet specific energy, 
and the mass load parameter, all measured at the Alfv\'en surface.
Each of the short lines shown in the plots is the result of a long-term evolution 
of a certain simulation (labeled with A, ..., H).
Each simulation has started from a different initial disk magnetization and has followed 
a different time evolution.

One way to correlate the jet kinematic properties to the underlying Keplerian disk properties is to assume that 
the energy and angular momentum at the disk surface are predominantly of kinematic origin 
\citep{2003ApJ...590L.107A}.
As a consequence, the $j$ and $\omega$ integrals (see Equations~\ref{eq:jint}~and~\ref{eq:wint})
do {\em not} depend on the underlying disk properties.
In other words, the relation between the jet energy and jet angular momentum is linear, 
see the definition Equation~(\ref{eq:jint}).

\begin{figure}
\centering
\includegraphics[width=9cm]{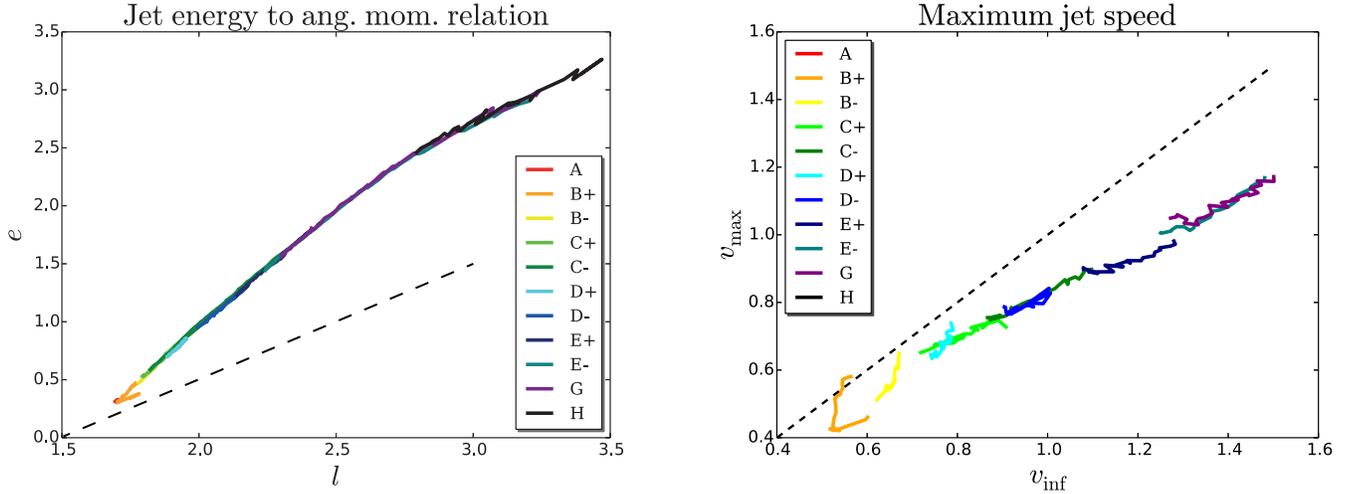}
\caption{Jet specific energy $e$ with respect to the jet specific angular momentum $l$. 
Each line represents the evolution of single a simulation (see Table~\ref{tbl:simuls}) 
from 700 to 10,000 time units.
The dashed line indicates the theoretical relation for cold jets $e = l - 3/2$.
}
\label{fig:jet_le}
\end{figure}

This is not true anymore if the underlying disk is strongly magnetized.
Figure \ref{fig:jet_mud} (upper left) shows the jet angular velocity $\omega$ with respect the disk 
magnetization. 
As we can clearly see, this physical variable now strongly depends on the disk magnetization
(this is also true for the kinematic integral $j$). 
Although both, the angular velocity $\omega$ and the kinematic integral $j$ of the jet are not constant, 
there is still a tight relation between the jet energy and the jet angular momentum
(Figure~\ref{fig:jet_le}). 
This relation is almost linear, but seem to follow two different linear regimes with a break
at about $l \simeq 2.7$.
It is interesting to compare this curve to the theoretical result for cold jets
$e = l - 3/2$ \citep{1992ApJ...394..117P, 2007prpl.conf..277P} that would correspond to a straight line 
below our numerically obtained relation shown in Figure~\ref{fig:jet_le}).
Thus, the jets launched in our simulations have more energy than theoretically
expected for cold jets.
This seems plausible as we do not consider cold jets but jets which carry some amount of enthalpy $h$ (in normalized units).
Therefore, a relation 
\begin{equation}
e = l - 3/2 + h 
\end{equation}
is expected between specific energy and specific angular momentum.
For $l$ \lax $2.7$ our curve follows a relation with $h \simeq e/2$, thus with a larger slope, while
for $l$ \gax $2.7$ our curve follows a relation with $h \simeq 4/3$, resulting in a slope similar to
cold jets, but offset by the enthalpy $h=4/3$.

Figure~\ref{fig:jet_mud} (upper right) shows the correlation between the actual disk magnetization and 
the total jet angular momentum.
Since the actual disk magnetization changes with evolutionary time, the corresponding 
jet angular momentum changes as well.
It can be noted that all simulations follow the same general trend - no matter at 
what evolutionary stage the actual simulation is.
A clear correlation between the disk magnetization and the jet angular momentum is established - the
higher the disk magnetization the more angular momentum is extracted by the outflow.

Similarly, Figure~\ref{fig:jet_mud} (lower left) shows the correlation between the jet energy $e$ and the disk 
magnetization $\mu_{\rm D}$.
We note that the magnetization is plotted logarithmically, while the energy is plotted in linear scale.
The curve shows a tight relation between the jet energy and the underlying disk 
magnetization - the stronger the disk magnetization $\mu_{\rm D}$ the higher the jet energy $e$.

We confirm the existence of the critical magnetization $\mu_{\rm crit} \approx 0.01$-$0.05$ found in paper I,
where the ejection-accretion process was examined.
Considering the change in the slope in Figure~\ref{fig:jet_mud} (lower left), also the correlation between the jet energy 
and the disk magnetization indicates two different regimes, separated at $\mu_{\rm crit}$.
As we have shown in paper I, below the critical magnetization the toroidal magnetic pressure gradient
plays a leading role in the outflow launching. 
The jets launched from this kind of configurations are usually known as {``}magnetic tower jets{''}.
For a magnetization above the critical value, the magneto-centrifugal acceleration is dominating.
The existence of these two regimes has been discussed by \citet{1997A&A...319..340F} considering
the outflow ejection efficiency indices. 
Low ejection efficiencies should lead to powerful centrifugally driven jets, high efficiencies to magnetic 
pressure-driven jets.
Here, we find a similar dichotomy, now in the framework of the disk magnetization.

The mass loading parameter $k$ (see Equations \ref{eq:kint} and \ref{eq:allint}) measures the amount of the matter that is 
ejected per unit magnetic flux.
Figure \ref{fig:jet_mud} (lower right) shows the correlation between the mass 
loading parameter $k$ and the disk midplane magnetization.
We see that for this MHD integral the different simulations converge not as tightly as for the integrals discussed before.

This is a result of the {\em gas density} present in the definition of $k$ (see 
Equation~\ref{eq:allint}).
Since the curves plotted in the figures show the continuous evolution of the disk-jet system, the disk and outflow 
densities are in principle different - within each simulation and even more between different simulations.
Considering these circumstances, the correlation shown in Figure \ref{fig:jet_mu_acej} (bottom)
is {\em remarkably} tight.

As pointed out by \citet{1997A&A...319..340F}, since cold jets can carry away the whole disk angular 
momentum, we may expect to find a systematic relation between mass loading and lever arm.
In fact, from our simulations we can confirm the existence of such a relation - an even stronger
relation than it was previously thought \citep{1997A&A...319..340F, 1997ApJ...482..712O}.
We find that for a disk magnetization lower than critical, $\MUdisk < \MUcrit$, the Alfv\'en lever arm follows 
a steep power law, $\lambda \propto k^{-5.8}$, while for $\MUdisk > \MUcrit$ the power law is flatter, 
$\lambda \propto k^{-2.3}$.
These relations simply follow from combining the correlation between $\lambda$ and $\mu_{\rm D}$ in 
Figure~\ref{fig:jet_mud} with that combining $k$ and $\mu_{\rm D}$.

Another useful quantity is the Alfv\'en lever arm, $\lambda = \ra / r_0$ 
(not shown).
It can be shown that it directly links the mass ejection and accretion rates \citep{1992ApJ...394..117P},
$\dot{M}_{\rm acc} \sim \lambda^2 \dot{M}_{\rm ej}$\footnote{Note that this is a simplistic approximation 
for our case and holds only for one field line. In general, both $\dot{M}_{\rm _acc}$ and $\dot{M}_{\rm ej}$ 
both depend on radius - as the accretion rate lowers when accreted matter becomes ejected from the disk.
}.
The jet angular momentum is related to the jet angular velocity through the 
Alfv\'en lever arm $l = \lambda^2 \omega$.
In Figure~\ref{fig:jet_mud} (bottom) we show a relation between the square of the lever arm 
$\lambda^2$ and the disk magnetization $\mu_{\rm D}$.
As for the jet energy and angular momentum, we see a strong relation of the Alfv\'en lever
arm with respect to the disk magnetization.
For a higher disk magnetization, the length of the Alfv\'en lever arm increases.
In case of a large Alfv\'en lever arm, more angular momentum is extracted from the disk.
Along with the angular momentum, also energy is extracted more efficiently.
These jets, carrying more angular momentum and energy, eventually become faster and wider.
Essentially, this is the regime of the Blandford-Payne magneto-centrifugal acceleration.

The clear correlations we found in our simulations provides a potential tool for understanding of 
the launching conditions of observed jets.
Knowing the jet energy $e$ would immediately provide us with the information of the actual magnetization in the 
disk that launches this outflow.
We note, that jet energy $e$ is a dimensionless quantity that incorporates the knowledge both
of the jet and its foot point (see Equation~\ref{eq:allint}).

The steady-state MHD conservation laws allow us, in principle, to connect 
the observed jet quantities far from the source to the launching conditions. 
\citet{1982MNRAS.199..883B} already provided analytical expressions for the 
energy and angular momentum partition for self-similar cold jets. 
Far from the source they find a ratio of Poynting flux to kinetic energy 
flux of $2/(M^2_{\rm FM}-2)$ for outflows with a fast magnetosonic Mach number $M_{\rm FM} > 1$. 
Applied to protostellar jets with typically $M_{\rm FM} \simeq 3$, 
this corresponds to jets that are kinematically dominated by a factor four.

Essentially, the jet terminal speed - defined by the jet terminal kinetic energy/momentum - depends 
on how much of the {\em magnetic} energy/momentum is left in comparison to the kinetic parts.
When the jet leaves the disk and enters the ideal MHD regime, the specific energy 
and specific angular momentum become fixed.
As the jet propagates, the magnetic energy and magnetic angular momentum are transferred 
into kinetic energy and kinetic angular momentum.
However, our simulation domain does not yet reach the asymptotic regime 
where energy and angular momentum conversion has converged.

\begin{figure}
\centering
\includegraphics[width=9cm]{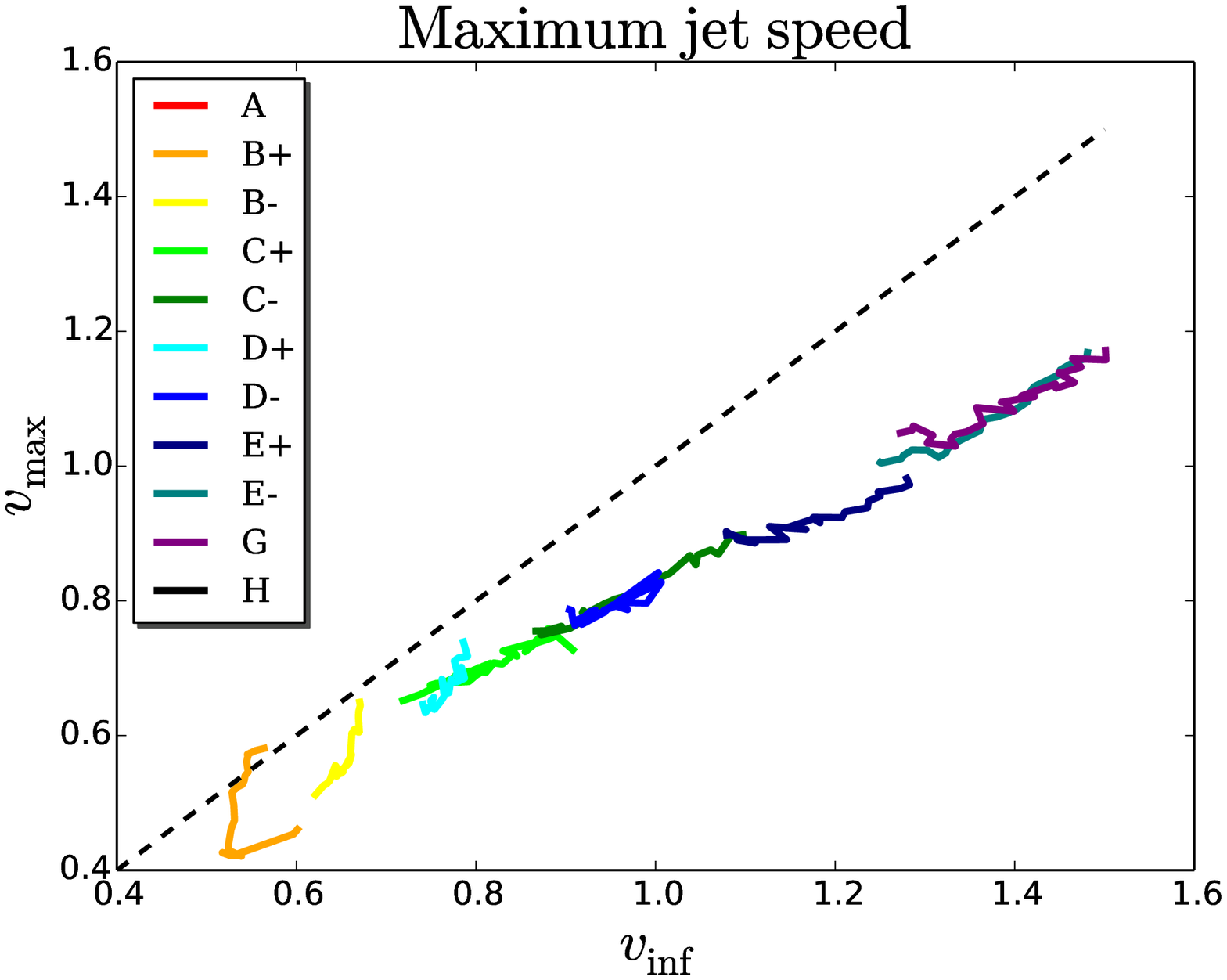}
\caption{Jet maximum speed $\vmax$ as measured in the domain with respect to 
the maximal jet speed estimated from the jet energy $e$.
Each line represents the evolution of a single simulation (see Table~\ref{tbl:simuls}) 
from 3,000 to 10,000 time units.}
\label{fig:jet_vinfvmax}
\end{figure}
 
In Figure~\ref{fig:jet_vinfvmax}, we show the maximum poloidal speed of the jet,
$v_{\rm max}=\max_z(\Vp)$, as defined by Equation~\ref{eq:vinf}, with respect to the theoretical 
limit for the jet velocity.
The velocity $v_{\rm max}$ is measured as the maximum speed of the jet in the whole domain
as soon as a quasi steady state of the simulation has been reached.
We find that in our simulations the theoretical limit is never reached.
The maximum speed obtained within the domain is about $75\%$ of the theoretical limit.

There may be several reasons for this.
First, the high-velocity jets obtained in simulations show a high degree of 
collimation. 
This is  different to e.g. self-similar studies, or studies
that assume a parabolic field distribution in general, both
implicitly assuming that the  magnetic flux diverges at infinity.
Therefore, some of the magnetic field energy is conserved along the 
flow as well as some gas enthalpy,
and not all energy is transformed into kinetic energy of the jet.

Another reason may be that energy conversion from magnetic to kinetic energy may indeed
take longer as the domain size allows.
Yet another reason may be the lack of resolution across the jet for very long distance in 
propagation direction (typical for a cylindrical jet on a spherical grid) - indeed
we found that an increase/decrease of the resolution across the jet leads to a higher/lower
jet maximum speed obtained in the domain.
While we can resolve the disk and the launching area with rather good resolution, the 
area of the collimated jet becomes less and less resolved along the direction of jet 
propagation. 
We suspect that this may lower the efficiency of energy conversion from magnetic to kinetic 
energy and also that the energy that is carried along the jet is somewhat lost, {``}diffused{''} 
away from the central spine.

\begin{figure}
\centering
\includegraphics[width=8.5cm]{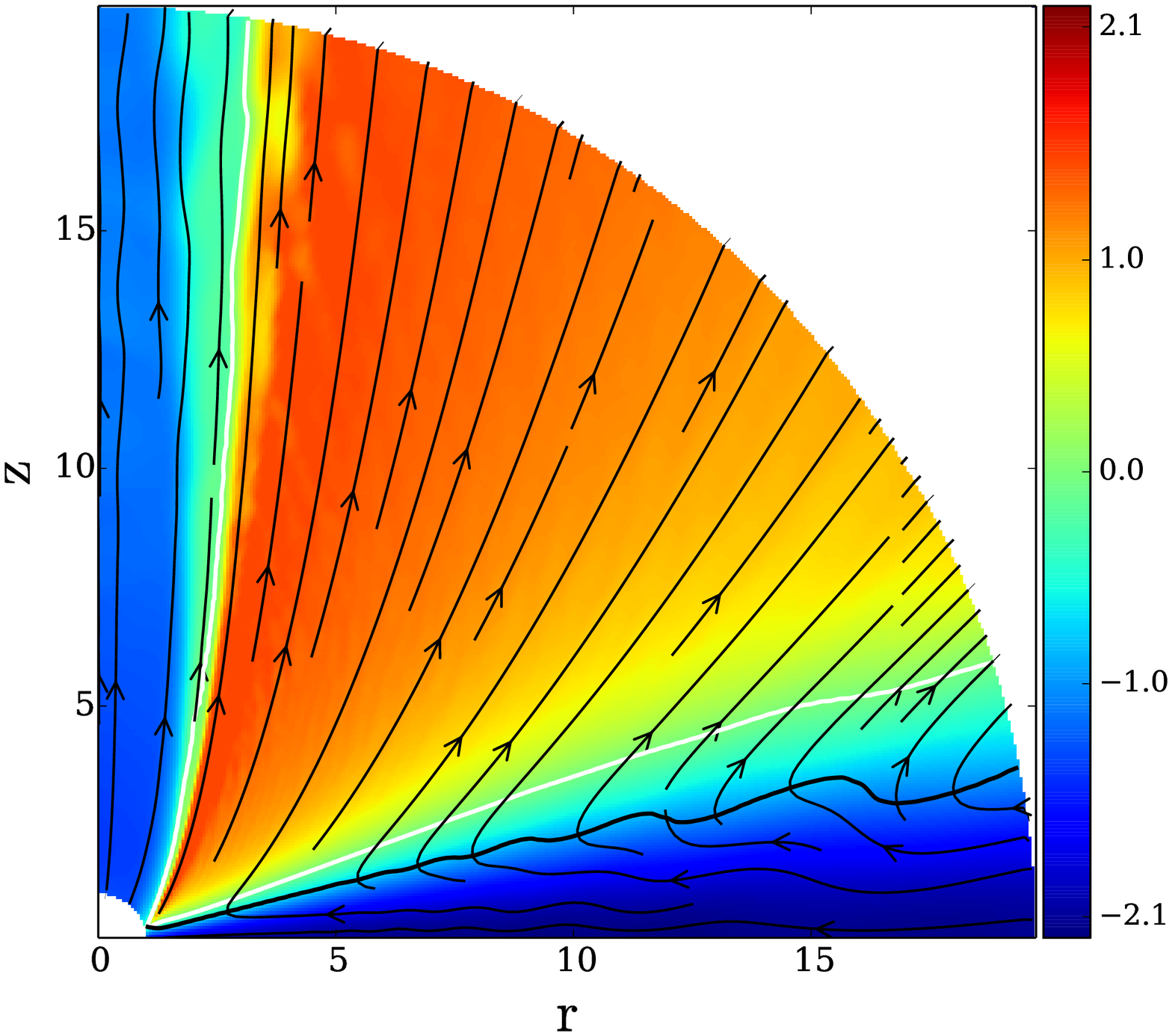}
\includegraphics[width=8.5cm]{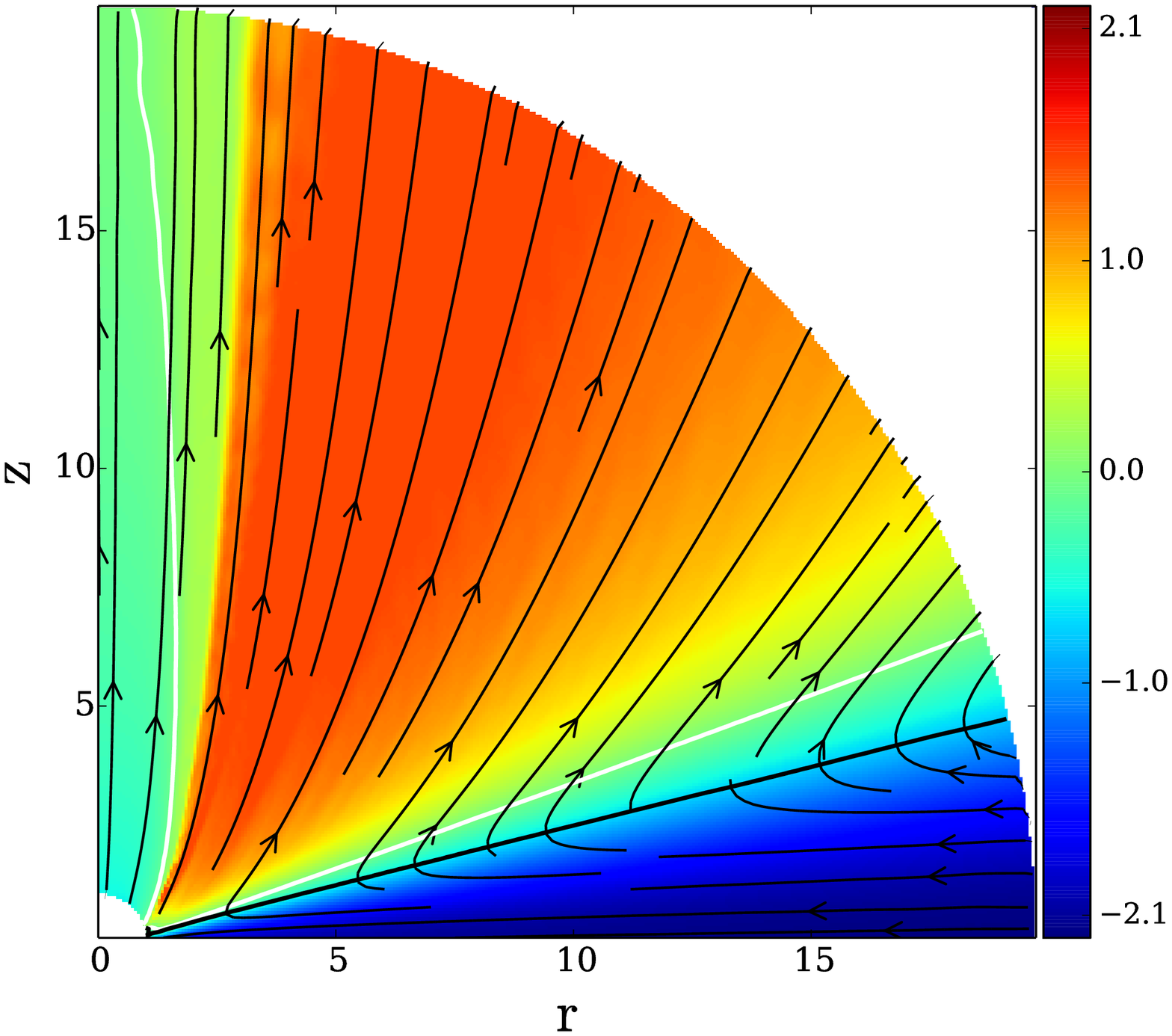}
\includegraphics[width=8.5cm]{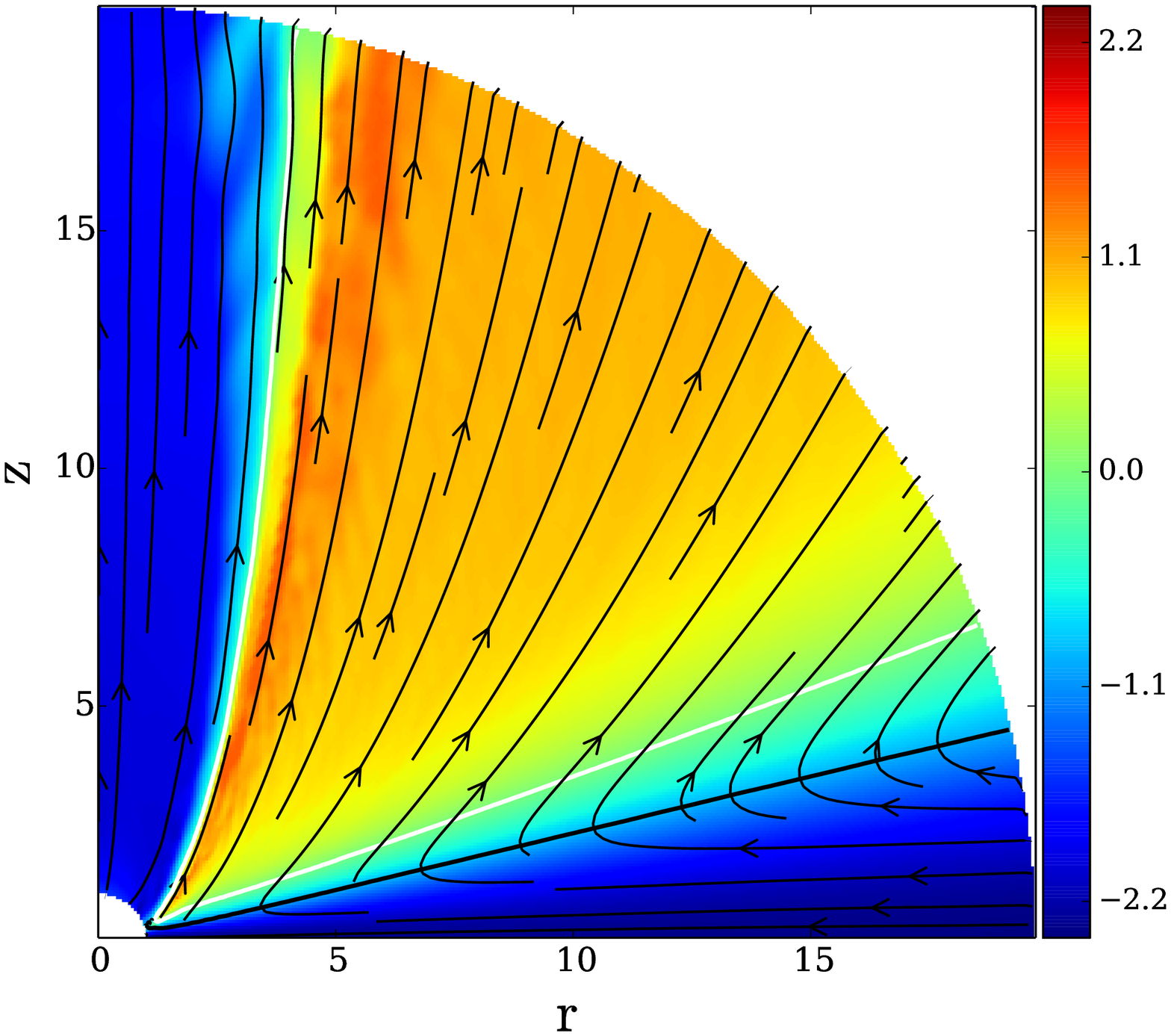}
\caption{Disk-jet magnetization $\mu (r,\theta)$
for simulation E+ at $t=1000$ (top) and $t=10000$ (middle), and for simulation D+ at $t=10000$ (bottom).
Shown is the 2D magnetization distribution $\log(\mu(r,\theta))$ (color coded), 
the surface where $\mu = 1.0$ (white line), 
the stream lines (thin lines with arrows), 
and the launching surface where $v_r = 0$ (black line). }
\label{fig:jet_mu_2D}
\end{figure}

\section{The accretion-ejection process}
In order to compute the mass, energy, and angular momentum fluxes,
we need to consider the correlations between the densities and velocities for both accretion and ejection.

\subsection{Mass loading and Alfv\'en lever arm}
\begin{figure}
\centering
\includegraphics[width=9cm]{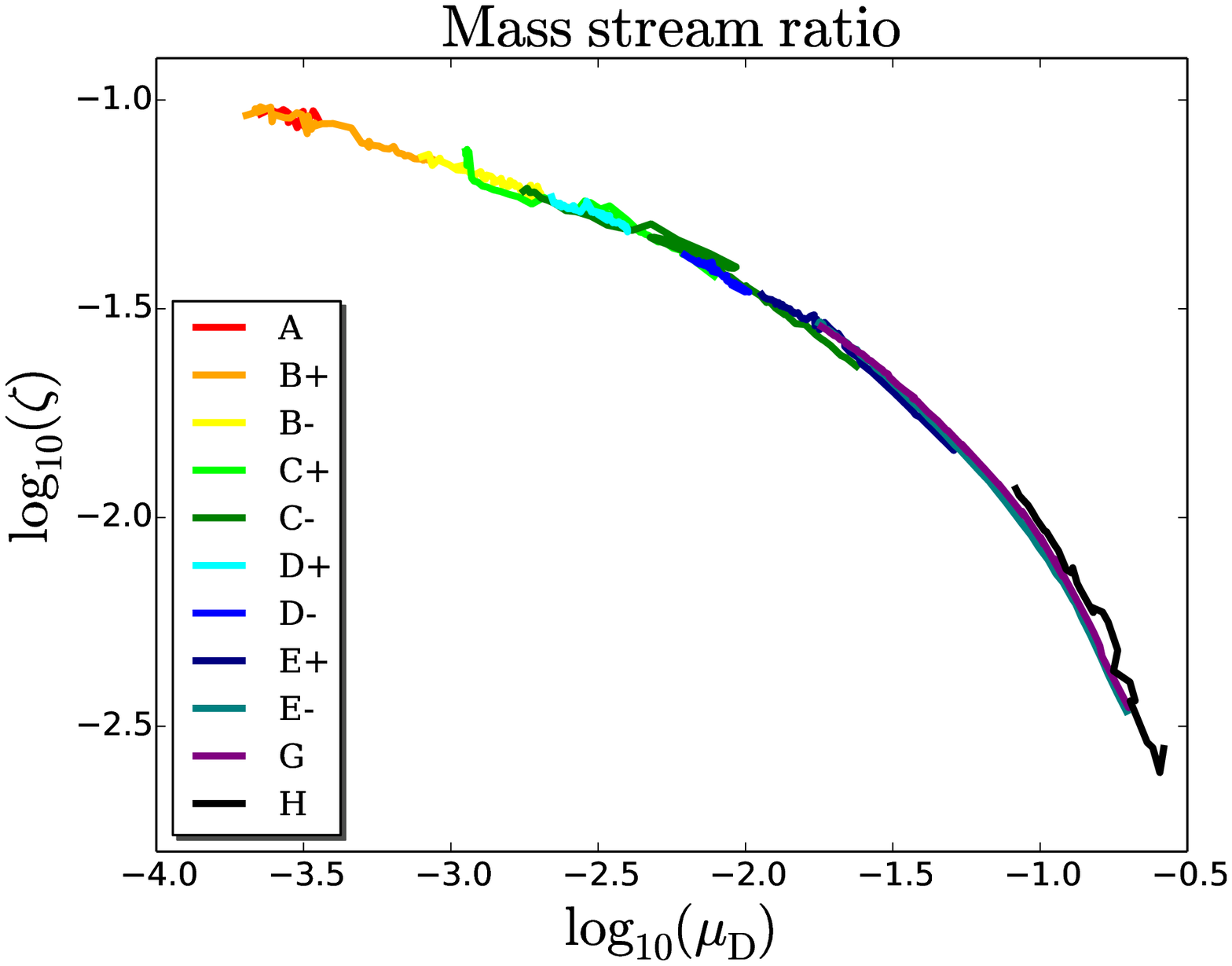}
\includegraphics[width=9cm]{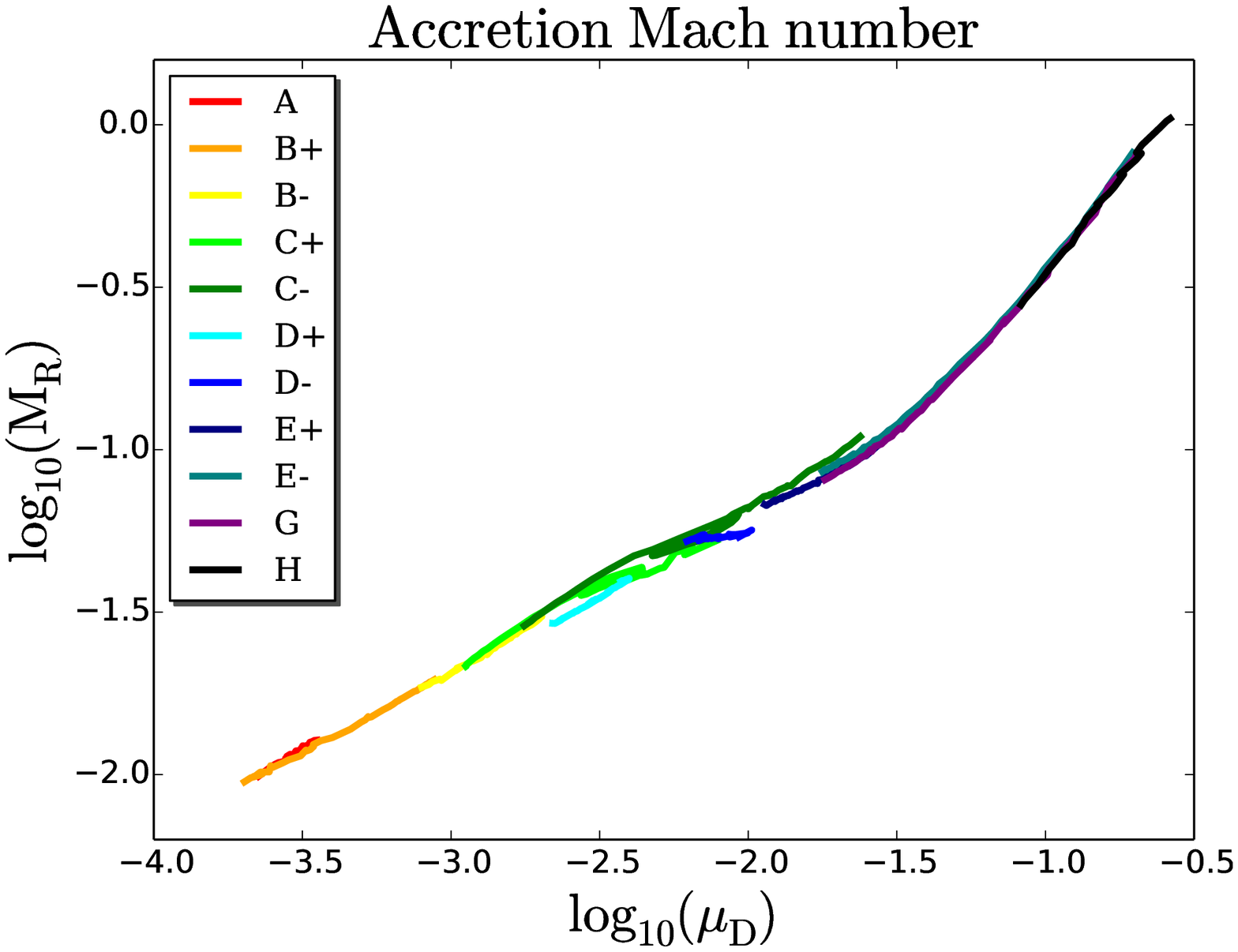}
\caption{Accretion-ejection properties with respect to the disk magnetization $\MUdisk$.
Shown is the mass stream ratio $\zeta$ (top) and
the accretion Mach number (bottom).
Each line represents the evolution of a single simulation (see Table~\ref{tbl:simuls}) from 700 to 10,000 time units.}
\label{fig:jet_mu_acej}
\end{figure}

It is helpful to introduce the mass stream ratio $\zeta$ that compares the properties of 
the ejected gas (i.e. the stream of the mass along the magnetic flux surface) at the Alfv\'en point to the 
value measured at the disk midplane,
\begin{equation}
\zeta = \frac{\rhoa \vpa}{\rhod \vpd}.
\end{equation}
Here, $\rho$ is the density and $\Vp$ the poloidal velocity at the disk midplane (subscript D) 
and Alfv\'en point (subscript A), respectively.
Note, that $\vpd = \Vr$, as the only poloidal component at the disk midplane is radial.
In Figure~\ref{fig:jet_mu_acej} (top) we show the mass stream ratio with respect to the disk magnetization.
We find a clear and very tight correlation between these quantities.
We note that this is a log-log plot indicating that the mass stream ratio is much
smaller for the case of strong disk magnetization.
In the case of weak and moderately strong magnetization, $\MUdisk < 0.1$, we find a 
relation $\zeta \propto \log ( \MUdisk )$ (not shown), thus a weak dependence of $\zeta$ from the disk magnetization.

It is known that there is also a tight correlation between the disk magnetization and the mean accretion Mach 
number $\Mr = \vpd/\Cs$ (see e.g. \citealt{1995A&A...295..807F, 2011ppcd.book..283K}, or paper I).
Figure~\ref{fig:jet_mu_acej} (middle) presents this relation for a wide range of the disk magnetization.
Our simulations confirm a tight relation between the mean accretion Mach number and the disk magnetization - also 
for the case of weakly magnetized disk.
The fact that the slope of the curves shown in Figure~\ref{fig:jet_mu_acej} (middle) changes, indicates that the accretion Mach 
number may grow at different rates when the the disk magnetization increases.
We again emphasize that the physical quantities plotted in Figure~\ref{fig:jet_mu_acej} (middle) do not belong to the same 
mass flow within one particular simulation, but they are obtained from different simulations.
Therefore, the relation we find is very general.
It is a consequence of the fact that the ejection and accretion processes are tightly connected.

\begin{figure}
\centering
\includegraphics[width=7cm]{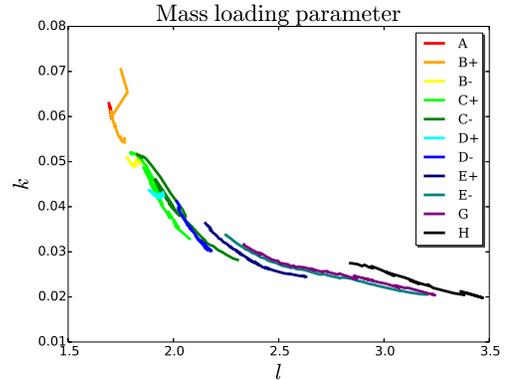}
\caption{Mass loading parameter $k$ and specific angular momentum $l$.
Each line represents the evolution of a single simulation (see Table~\ref{tbl:simuls}) from 
700 to 10,000 time units.}
\label{fig:jet_lk1}
\end{figure}

Figure~\ref{fig:jet_lk1} shows the mass loading parameter $k$ with respect to the jet angular momentum $l$.
This presentation is similar to the classic diagram presented in \citet{1982MNRAS.199..883B} and \citet{2000A&A...353.1115C}
who plot the correlation
between lever arm and mass loading\footnote{Note our slightly different notation. Our definition of $\lambda^2$ 
corresponds to their $\lambda \equiv \left(\ra / r_0 \right)^2$
\citep{2000A&A...353.1115C}, while our $k$ is similar to their $\kappa$.}.
We have plotted $k(l)$ and not $k(\lambda)$, however, since $l=\lambda^2 \omega$, and $\omega \simeq 0.9$, the 
presentations are directly comparable to the literature cited above.
Note that these figures shows the correlation between two jet properties - an correlation as 
already expected from the steady MHD theory while Figure~\ref{fig:jet_mu_acej} (bottom) essentially 
{\em correlates jet properties to the underlying disk properties}.
We first note that our curves follow the same trend as for the literature papers -
an increase in $k$ by a factor of 3 implies an decrease in $l \simeq \lambda^2$ by a factor 2 in our case, 
while for \citet{2000A&A...353.1115C} we find 
an increase in their $\kappa$ by a factor of 4 with a corresponding decrease in $\lambda$ 
(our $\lambda^2$) by a factor 2 (for their $\omega_A = 1.3$).

We interpret this as a general agreement concerning the {\em physics} of the mass loading process.
However, there is a strong difference in the number values for the lever arm.
While in our case $l \simeq \lambda^2 \simeq 2 ... 3$, \citet{1982MNRAS.199..883B} and \citet{2000A&A...353.1115C} 
find lever arms of typically 7...10.
We interpret this difference as a difference of the {\em geometry} of the accretion-ejection structure.
In general, the lever arm found in our simulations are smaller than those derived from analytic models.
In the simulations, the outflow ejected from the inner disk seem to either become collimated so fast that 
the Alfv\'en point along the magnetic field line is not far in radius from the foot point of the field line,
or the Alfv\'en surface is located close to the launching surface of the flow that has not yet expanded 
when it becomes Alfv\'enic. 
An example of the latter can be seen in Murphy et al.~(2010) (Fig.~1).
To our knowledge this is a general behavior found in simulations of jet formation (independent whether they include
disk structure or not in their simulations).
Whether this effect is due to a principle difference in the geometry of the system, and maybe related to the
self-similar approach of the steady-state solutions, is an essential question of jet launching and needs to
be addressed in a future study.

In order to disentangle the different mechanisms affecting the ejection process, we rewrite the mass loading 
parameter $k$ using the quantities introduced and discussed above,
\begin{equation}
k = \frac{\sqrt{\gamma} \Mr \zeta \lambda^2}{\sqrt{2\mu}},
\end{equation}
where we applied $\bpa = \bpd / \lambda^2$.
This equation simply follows from substituting $\Mr = \vpd/\Cs$ and $\bpd/\sqrt{\rhod}=\sqrt{2\mu}\CsT$
into Equation~\ref{eq:kint}~and~\ref{eq:allint}.

We can now compare how the different variables in the mass loading parameter change with the disk 
magnetization, and, thus affect $k$.
As we have previously discussed, in the case of the moderately magnetized disk, the accretion Mach number 
increases almost linearly with the disk magnetization (see above and paper I).
From Figure~\ref{fig:jet_mud} (lower right) it becomes obvious that the Alfv\'en lever arm squared
$\lambda^2$ also increases with the disk magnetization.
However, this increase is rather weak - a change in the disk magnetization by three orders of magnitude leads only
to a factor two in the change of $\lambda^2$.
We have, however, found above that the mass loading parameter $k$, {\em decreases} with the disk 
magnetization.
One possible explanation for this behavior is that the ejection to accretion stream ratio, $\zeta$, decreases 
very quickly with the magnetization, as we have shown in Figure~\ref{fig:jet_mu_acej} (top).
Note that analytical models are able to provide an exact link between the lever arm $\lambda$,
the mass load parameter $k$, and a parameter that quantifies the so-called ejection efficiency 
$\xi$ \citep{1997A&A...319..340F, 2000A&A...353.1115C}.

\subsection{Jet launching and acceleration}
In this section, we finally discuss the difference in the jet-launching mechanism for different disk magnetizations.
In paper~I we have found indication for a critical value for the disk magnetization $\MUcrit \approx 0.03$ that seem to 
separate two regimes of disk dynamics.
These two regimes can be referred to as advection or diffusion dominated, respectively.

Here, we would like to confirm this finding again by considering the magnetic Reynolds number for accretion
$Re_{\rm m} = {\Vr H}/{\eta}$, 
where $\Vr = 4 \MUdisk \Cs$ approximates the accretion velocity and
$\eta = \am \sqrt{\MUdisk} \Cs H$ corresponds to the magnetic diffusivity prescription applied.
Thus, the resulting accretion Reynolds number depends mainly on the disk magnetization $\mu$ 
and the diffusivity parameter $\am$
\begin{equation}\label{eq:rey}
Re_{\rm m} \approx 2\sqrt{2} \frac{\MUdisk}{\am}.
\end{equation}
However, this is an approximation and serves mainly to illustrate the accretion-diffusion balance.
We found that the diffusivity parameter $\am \sim 1.6$  
yields a quasi-steady state regime for accretion such that diffusion and advection are roughly in 
balance (thus $Re_{\rm m} \sim 1$).
In other words, the same amount of the magnetic flux carried inward by the accretion is diffused outward.
We note that the aforementioned reasoning can be only qualitative, as the physical evolution of the system
is highly complex.

By examining the properties of the jets launched from the disks, we can 
distinguish two different, but complementary regimes of the jet launching.
This can be seen, for example, in Figure~\ref{fig:jet_mud} (upper left), where we find a prominent peak in the jet angular 
velocity around $\MUdisk \approx 0.01-0.05$.
In order to better understand what is the difference between these regimes, we now present several jet quantities 
computed at the Alfv\'en point.

\begin{figure}
\centering
\includegraphics[width=9cm]{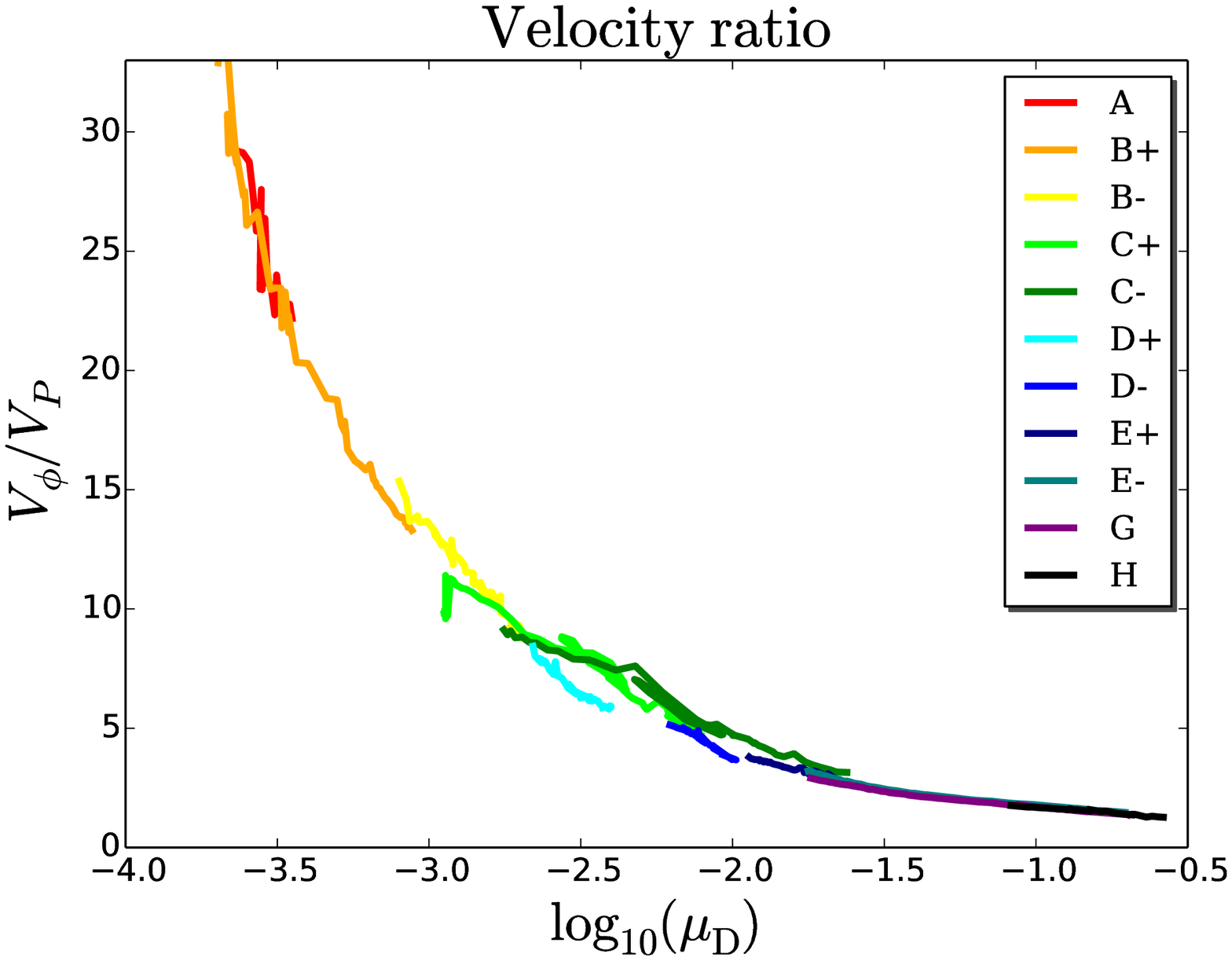}
\includegraphics[width=9cm]{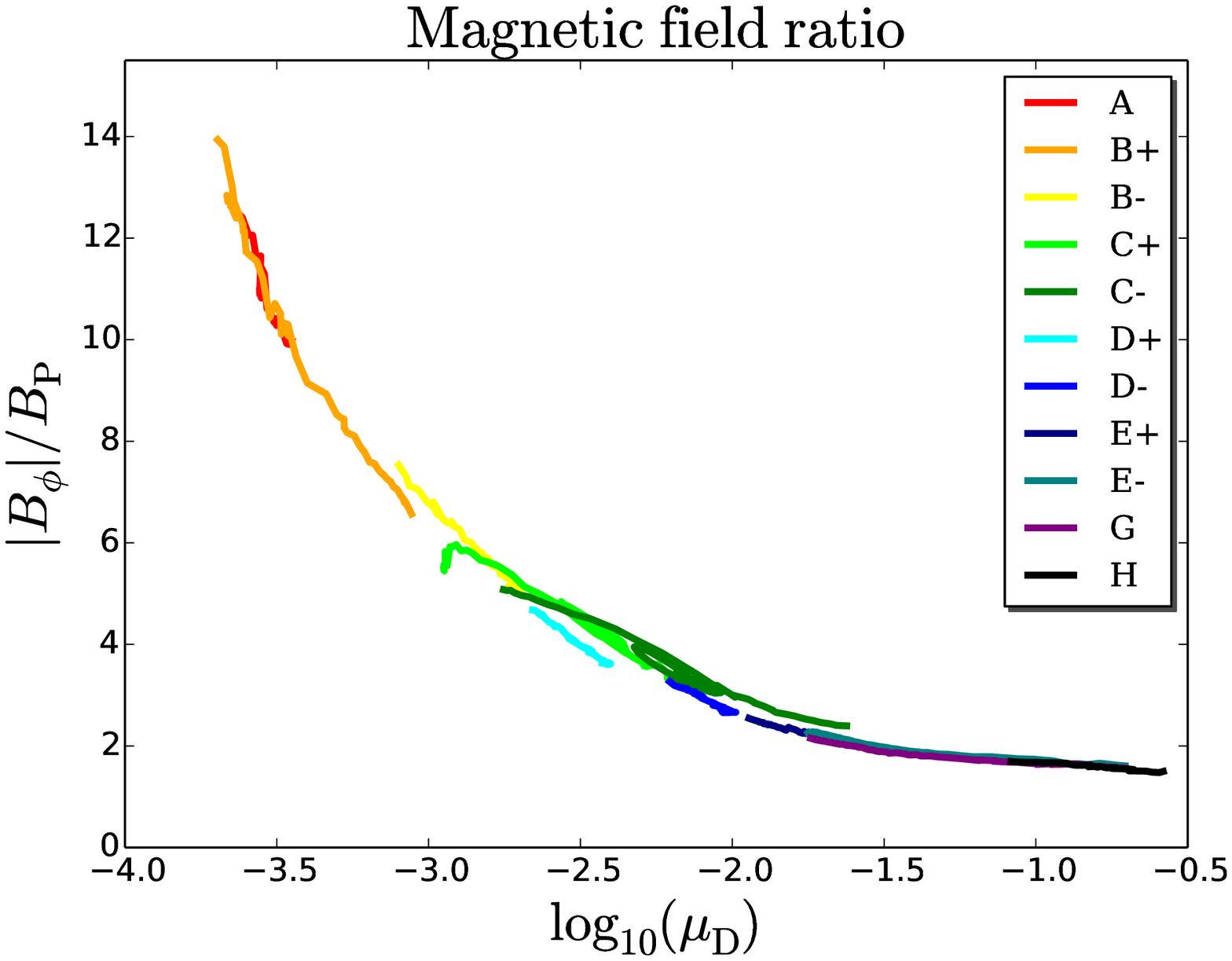}
\caption{Ratio of the toroidal to poloidal components of the velocity (top) and the
magnetic field (bottom), measured at the Alfv\'en point, with respect to the disk magnetization $\MUdisk$. 
Each line represents the evolution of a single simulation
(see Table~\ref{tbl:simuls}) from 700 to 10,000 time units.}
\label{fig:jet_mu_v_b_x}
\end{figure}

Figure \ref{fig:jet_mu_v_b_x} shows the ratio of toroidal to poloidal velocity (top) and magnetic 
field (bottom), computed at the Alfv\'en point.
As the disk magnetization decreases, the poloidal velocity decreases as well, leading to an increase of the 
toroidal to poloidal velocity ratio.
For a low disk magnetization, the Alfv\'en surface is located rather close to the disk surface.
Because of the small poloidal velocity, the induction of the toroidal magnetic field (by the toroidal shear)
is rather efficient, leading to a high toroidal to poloidal magnetic field ratio.

The excess of the magnetic energy stored in the toroidal magnetic field component, together with the smaller angular 
momentum loss rate result in an outflow that is driven predominantly by the toroidal magnetic field pressure 
gradient, and not magneto-centrifugally. 
This regime of jet launching is usually known as producing {``}tower jets{''}.
In the case of strong disk magnetization, the poloidal field dominates the toroidal magnetic field, while the 
extraction of disk angular momentum extraction is more efficient. 
As a consequence, the Alfv\'en lever arm is larger, and the mass loading is lower.
This is the regime is of the Blandford-Payne magneto-centrifugal launching.

The transition between the two regimes may appear smooth.
Naturally, further details may depend on the diffusivity prescription that is applied.
The ability of the disk to launch and sustain {\em tower jets} relies on its ability to
generate a strong {\em toroidal} magnetic field component.
However, in the standard approach, the diffusivity is parameterized only by the poloidal component of the 
magnetic field.
On the other hand, a magnetic diffusivity prescription that considers both magnetic field components (as in
\citealt{2014ApJ...796...29S}) is able to suppress some of the launching regimes.

It is interesting to compare Fig.~\ref{fig:jet_mu_v_b_x} to previous studies.
\citet{2000A&A...353.1115C} have found in self-similar studies of jet launching that
the quantities plotted in Fig.~\ref{fig:jet_mu_v_b_x} follow certain interrelations.
That is 
\begin{equation}
\left[\frac{B_{\phi}}{B_{\rm P}} \right]_{\rm A} =     - g_{\rm A}   \omega_{\rm A}\,,\quad
\left[\frac{v_{\phi}}{v_{\rm P}} \right]_{\rm A} =  (1 - g_{\rm A})  \omega_{\rm A}\,,
\label{eq:fastness}
\end{equation}
where $g \equiv 1 - \Omega/\Omega_{\rm F} $ and the so-called fastness parameter 
$ \omega_{\rm A} = \Omega_{\rm F} r_{\rm A} / V_{\rm Ap, A}$ with the poloidal 
Alfv\'en speed at the Alfv\'en point $r_{\rm A}$.
For a low magnetization $\mu_{\rm D} \simeq 10^{-3.5}$ we derive from our data $g_{\rm A} \simeq 1/3$ and $\omega_{\rm A} \simeq 36$, while
for a high magnetization $\mu_{\rm D} \simeq 10^{-1}$ we derive from our data $g_{\rm A} \simeq 1/3.5$ and $\omega_{\rm A} \simeq 14$.
These number values, derived from our simulations (Fig.~\ref{fig:jet_mu_v_b_x}) and using the equations \ref{eq:fastness} are 
consistent with a direct estimate of the fastness parameter, i.e.  applying $\Omega_{\rm F} \simeq 0.9$ at the foot point of the
field line, and an Alfv\'en radius of $r \simeq 1.8 - 2.6$, resulting in $\omega_{\rm A} \simeq 15-20$.
Note that these values of the fastness parameter are on a different parameter range compared to the case of
the self-similar studies resulting in $\omega_{\rm A} \simeq 1-2$ \citet{2000A&A...353.1115C}.
We believe that this is mainly due to the fact that our study is {\em not} self-similar. 
The flux surfaces we investigate are the most innermost ones routed in the disk, thus
deviating most from a self-similar structure.
Also, collimation in our non-self-similar case is more rapid, resulting in a relatively small lever arm and an Alfv\'en 
radius only at $r \simeq 1.8 - 2.6$. 
Since the strong collimation, the acceleration is less efficient, resulting in a rather low velocity at the Alfv\'en point.
Also the asymptotic speed along this flux surface is lower than expected from self-similar studies.

Having above mainly discussed the {\em jet acceleration} mechanism, a remaining question is the actual {\em launching} 
mechanism of the outflow.
It has been well demonstrated \citep{1997A&A...319..340F,2007A&A...469..811Z, 2010A&A...512A..82M, 2012ApJ...757...65S}
that the main 
driving force for lifting the disk material out of the disk into the outflow and thereby passing the sonic surface is 
the gas pressure gradient.
The poloidal Lorentz force is negative, thus does suppress the vertical lifting of material below several disk pressure 
scale heights.
We note that the jet material is not lifted directly from the disk midplane into the outflow.
It is the material close to the disk surface that is lifted in the outflow \citep{1997A&A...319..340F, 2012ApJ...757...65S}.

Note that we need to distinguish between the {\em launching} and the {\em acceleration} mechanism.
While disk winds seem to be launched in general by the gas pressure gradient, the outflow
acceleration mechanism is different for weakly or strongly magnetized outflow, respectively.
As discussed above, weakly magnetized jets are mainly driven by the magnetic pressure gradient,
while highly magnetized jets are mainly driven by the magneto-centrifugal acceleration
(see also \citealt{2012ApJ...757...65S}, where jets of different plasma beta and diffusivity 
were studied).

We should point out a potential numerical bias for the launching mechanism.
As discussed in \citet{2010A&A...512A..82M}, the strong gradient in density across the disk surface
(note the initially Gaussian density and pressure profile in vertical direction) leads to a 
numerical flux of matter.
This is in addition to the mass loading that is allowed by the physical magnetic diffusivity and
is dependent on the forces in the surface layer.
Since our numerical resolution is higher than in the Murphy et al. paper\footnote{
Murphy et al. apply a resolution of 1.3 cells per disk scale height $h$ at the inner disk radius, 
assuming that $h = 0.1 r$.
For larger radii the resolution increases to e.g. 13 grid cells per $h$ at $r=10$.
We apply a resolution of 8 grid cells per scale height $H = \epsilon R$ for all disk radii.
},
we expect the numerical mass loading to be weaker, but still present.
Note that we also observe a jump in entropy across the disk surface as discussed in \citet{2010A&A...512A..82M}.
Due to our higher resolution this effect of numerical heating is smaller, but still present.


\subsection{Comparison to other studies}
We have already mentioned the works of \citet{2009MNRAS.400..820T} and \citet{2010A&A...512A..82M} 
who were first in studying the jet launching in particular with respect to the disk magnetization.
We now want to compare our results to the key findings of \citet{2010A&A...512A..82M}. 
Our simulation time scale seems to be somewhat larger\footnote{
We mention a discrepancy within the \citet{2010A&A...512A..82M} paper concerning
the evolutionary time of their simulations.
We adopt a time unit $T_0 \equiv R_0 / V_{\rm K,0}$ that
arises naturally in the dimensionless of MHD equations
(see e.g. \citealt{1997ApJ...482..712O, 2002A&A...395.1045F, 2007A&A...469..811Z, 
2009MNRAS.400..820T, 2013MNRAS.428.3151T, 2012ApJ...757...65S}),
and use as length unit the inner disk radius $R_0$ and as velocity unit the Keplerian velocity at $R_0$
(as also Murphy et al. do).
Murphy et al. apply a vertical grid extension of $840 r_0$ claiming that this 
{``}{\em ensures that the magnetized outflow never reaches the boundaries}{''} (quotation).
However, they also state simulation times of 950 in their (orbital) time units, 
thus $2\pi 950 T_0 \simeq 6000 T_0$.
From their Fig.~4 we derive a jet speed $V_{\rm jet} \simeq V_{\rm K,0} = 1.0$.
In {\em their} time units the jet crosses the whole grid by
$T_{\rm cross} = 840 R_0 / V_{\rm K,0} = 840 T_0 = 134 \times 2 \pi T_0$.
Within 950 of {\em their} time units the jet would propagate $6000 R_0$, well beyond the grid 
boundary.
Thus, there seems to be an inconsistency either in their claim of a grid boundary such far
that the outflow does not reach it or in their claim of simulation run time.
As the authors do not show a time series on the global grid, the outflow evolution cannot 
compared directly with other simulations.
}, 
however we know from our much longer-lasting simulations (up 500,000 dynamical time steps) 
that the evolutionary state we investigate is indeed in steady state.

Arguably the most essential result from \citet{2010A&A...512A..82M} is the finding that
super-fast jets can be launched only from the very central disk area where the magnetization
is strong enough.
This is demonstrated in their Fig.~15 that shows the radial disk and jet magnetization profile
along the launching surface\footnote{We feel obliged to mention an inconsistency 
concerning this figure. While the authors define the disk surface in Sect.~3.1 in their paper 
as the surface where $V_r$ changes sign (as we do), 
in their Fig 15 they plot $\sigma^{+}$ that is defined in equation 29 in their paper at $z=h$
(however, after equation 29 they state that $\sigma^{+}$ is measured at the disk surface,
implying, in contradiction to their earlier definition, that $z=h$ is the disk surface). 
}
and claim that outside of about 5 inner disk radii the magnetization is too low to launch fast 
jets\footnote{We note that a proper radial scale for the horizontal axis is not really provided, 
as only one label is shown for the radius tick-marks, and no inner radius for the plot is
stated. Also it remains unclear whether a linear or log scale is applied for the horizontal axis.
}.
Let's assume that the magnetization profile in their Fig.~15 is actually plotted for the disk 
surface defined by $V_r$ changing sign and the size of the jet launching area of the disk
is indeed 5 inner disk radii.
We now discuss our findings in the light of their result. 
In Figure~\ref{fig:jet_mu_2D} we show the 2D profiles of the disk-jet magnetization of our
simulation E+ at times $t=1000$ and $t=10000$, and likewise for simulation D+ at $t=10000$.
Simulation E+ starts with a magnetization parameter $\mu_0 = 0.01$. 
At $t=1000$ and $t=10,000$ the actual disk magnetization in the area between $r=1.1$ and $r=1.5$ 
(the launching area of the fast jet), has reached values of $mu_{\rm D} = 0.012$ and
$\mu_{\rm D} = 0.049$, respectively\footnote{These two snapshots in time are therefore seen on different
positions along the dark blue tracks denoting the evolution of run E+ in 
Figures \ref{fig:jet_mud}, \ref{fig:jet_mu_acej}, \ref{fig:jet_mu_v_b_x}.}.
Simulation D+ is initially weaker magnetized, characterized by $\mu_0 = 0.005$ and reaches a disk 
magnetization of only $\mu_{\rm D} = 0.00218$ at $t=10,000$ in the same launching area.

In Figure~\ref{fig:jet_mu_2D} we have further indicated the surface of jet launching, here defined as the 
surface where the poloidal velocity changes from accretion to ejection, thus where $v_r$ changes sign.
This surface where $v_r =0$ is indicated in the panels by a black line.
We find that this surface is typically located along $z(r) = 2 H(r)$ in our case, thus two times higher 
than indicated by the equation 29 in \citet{2010A&A...512A..82M}, but similar to Fig.~1 in \citet{2010A&A...512A..82M}.
In our figure we have also indicated the surface where $\mu = 1.0$ (white line).
This location is typically at altitudes somewhat higher then the launching surface, at $z(r) \simeq 3 H(r)$. 

Note that these surfaces follow a radial path throughout the panels in Fig.~\ref{fig:jet_mu_2D}, that
are subsets of the whole simulation domain extending till $R=1,500$ at $t=10,000$.
These subset also show an area of simulation that has definitely reach a quasi-steady state.
Here is a maybe another difference to the Murphy et al. paper, from which it is not clear to us,
whether those simulations have reached a quasi-steady state.

In our simulations we therfore observe the following situation.
Jet launching - the transition from accretion to ejection - may happen for lower values of disk 
magnetization compared to \citet{2010A&A...512A..82M} - in general the outflow is launched from
a region area where $\mu(r,z) \simeq 0.1$.
The magnetization increases rapidly with vertical height providing the energy reservoir for 
subsequent jet acceleration.
We find an increase of magnetization by a factor of ten within 
$\Delta z \simeq 1.0$ (from $\mu = 0.1$ to $\mu =1.0$ from $z(r)=2H(r)$ to $z = 3H(r)$.

This configuration in the jet launching region is general and does not alter substantially for the 
simulation runs we have investigated.
However, we find that the simulations with higher disk magnetization also have a higher initial jet 
magnetization (see the disk corona magnetization in the middle and lower panels in Fig.~\ref{fig:jet_mu_2D}),
and thus generate a more energetic jet.
We mention that we find similar results for a slightly altered magnetic diffusivity profile, 
although this issue deserves a closer investigation in the future. 
In \citet{2012ApJ...757...65S} we have thoroughly investigated the impact of the diffusivity profile 
and magnitude on the launching process.

We now come back to the finding of the Murphy et al. of a {\em radially} limited disk area for jet launching, 
derived from the local disk magnetization.
This result was essentially obtained by evaluating the {\em radial disk magnetization profile} of one simulation 
run for a low disk magnetization.
In our paper, however, we investigate properties of jets launched from the innermost disk, considering 
a number of simulation runs resulting in a broad range of disk magnetization {\em in that area}.
On the other hand, our results may be further extrapolated such, that if outflows are launched from different 
radii along the disk, the properties of these outflow layers change with the {\em local} disk magnetization 
at  the outflow footpoint - as described by the interrelations we have found.
In contrary to Murphy et al. we do not find a locally constrained disk area for jet launching.
This is partly by purpose.
In order to reach very long simulation times, for this paper we have chosen a certain setup for the
disk magnetization - the so-called strong diffusivity model - developed previously by \citet{2014ApJ...793...31S},
that allows for a smooth and steady accretion for very long simulation times.
In this approach the advection and diffusion of magnetic flux leads to a steady-state {\em radial} magnetization 
profile that is roughly constant (see Fig.~23 in \citealt{2014ApJ...793...31S}).
This is comparable to the magnetization profile of \citealt{2009MNRAS.400..820T}), who impose a constant
magnetization profile along the equator.
Thus, an outflow is launched from all over the disk surface in this model.

Note that discussion above again underlines the importance of the magnetic diffusivity model that is not know 
from first principles (as probably provided by global disk models considering direct simulations of the
MRI including also further physical effects such as resistivity or heating/cooling).
Still, we think that we may safely accept general conclusions such as from Murphy et al. concerning the 
limited disk area of jet launching, 
or the interrelation derived in the present paper between the disk magnetization and the jet 
properties.
As long as we have to make a choice about the magnetic diffusivity distribution applied,
the exact number values, such as the area of jet launching or e.g. the actual value of mass loading will still
depend on the disk diffusivity model.

We therefore consider both approaches - the one presented here and the one by \citet{2010A&A...512A..82M} - as 
complementary.

The work by \citet{2009MNRAS.400..820T} is in some sense similar to ours as imposing a constant magnetization
along the disk mid plane initially. 
Thus, the values for the magnetization used should be directly comparable to our $\mu_{\rm D}$.
However, they do not discuss the time evolution of the magnetization, nor the vertical profile of it.
The range of disk magnetizations applied is from 0.1 to 3.0, thus smaller, although they mention a highly
unsteady evolution for two very low magnetization levels, $\mu = 0.01$ and $\mu = 0.001$, with maximum
wind velocities of 20\% of the Keplerian speed at the footpoint.
In our simulations we find correspondingly low outflows speeds although higher by a factor 3 for $\mu = 0.001$.

\section{Conclusions}
We have presented results of MHD simulations investigating the launching of jets from
magnetically diffusive accretion disks.

In a previous paper \citep{2014ApJ...793...31S} we have presented test simulations of our setup in 
spherical coordinates.
We had further demonstrated that the disk accretion, ejection and, in particular, the accretion-to-ejection 
ratio of the mass and energy fluxes strongly depend on the actual disk magnetization.
In the present paper we have applied a novel approach and have studied the relation
between the jet properties,  i.e. mainly the steady state MHD integrals, and the 
underlying disk properties, with emphasis on the disk magnetization $\mu_{\rm D}$.

Our work extends the results previously published by \citet{2009MNRAS.400..820T} and \citet{2010A&A...512A..82M}
by investigating a much larger parameter space for the disk magnetization.

While \citet{2010A&A...512A..82M} investigate how jet launching is affected by the
magnetization profile along the disk surface,
we follow a complementary approach and concentrate on the innermost disk area investigating 
the properties of jets launched from there considering a broad range of actual disk magnetization 
of $\mu_{\rm D} = 10^{-3.5} ... 10^{-0.7}$.
The disk magnetization at this jet launching area follows from a simulation of the global accretion-ejection
dynamics, and considers a quasi steady state.

As our main result, in this paper we present a series of correlations between the actual 
values of the jet conserved MHD quantities, 
such as energy, angular momentum, lever arm, or jet rotation, and the disk magnetization.
These correlations cover more than three orders of magnitude of the disk magnetization.

Although the actual number values of the physical quantities presented in our study 
may depend on further details of various model prescriptions 
(such as e.g. the magnetic diffusivity that itself is a result of the disk microphysics 
causing the turbulence),
we believe that the relations we have obtained are robust in general, at least qualitatively.
In particular, we have obtained the following results:

(1)
We have established a set of correlations between the dynamical properties of the jet and the 
underlying accretion disk.
In particular, we have shown how the four jet MHD integrals are connected to the underlying 
disk magnetization.

(2)
In particular, a high disk magnetization results in \\
i) a larger Alfv\'en lever arm of the jet,\\
ii) a larger jet specific angular momentum and energy,\\
iii) a lower mass loading parameter $k$,\\
iv) a lower mass ejection to accretion ratio,\\
v) a higher accretion Mach number,\\
vi) a lower velocity and a lower ratio of the toroidal to poloidal magnetic field component at the Alfv\'en surface,\\
vii) a higher asymptotic jet speed,\\
viii) a lower jet angular velocity.\\
For a  comparatively low disk magnetization, the opposite of i)-vii) is true,
while jet angular velocity also decreases, 
having a peak at a magnetization that is close to the critical magnetization 
that is about $0.01-0.05$ for our setup.

(3)
While \citet{2010A&A...512A..82M} previous papers find indication for a limited size of the jet 
launching area, as the jet and disk magnetization falls below a critical value at a certain radius,
our simulation give a different result. 
For the wide range of disk magnetization we investigated for the inner disk area, we find outflows
for all these magnetizations.

(4)
The launching of the outflow, thus lifting the accreting material into the outflow is done by the
disk vertical pressure gradient, in agreement with previous studies
\citep{2009MNRAS.400..820T,2010A&A...512A..82M,2012ApJ...757...65S}.

(5)
Having defined the disk surface as the surface where accretion turns into ejection, thus $v_r =0$, 
we find that it is located roughly at two disk thermal scale heights $z(r) \simeq 2 H_{\rm T}(r)$.
At this position the magnetization is somewhat below unity.
The magnetization increases rapidly above the disk surface, providing the energy source for jet
acceleration.

(6) 
In agreement with our previous studies (see e.g. \citealt{2012ApJ...757...65S}), 
the correlations we find between the jet parameters and the disk magnetization suggest the existence 
of two acceleration mechanisms at work.
The correlations we find between the jet dynamical parameters and the disk magnetization typically 
show a different slope for low and high disk magnetization.
Also the relative magnetic field strength of the toroidal component increases with decreasing
disk magnetization.
We interpret that acceleration is predominantly by magnetic pressure gradient for weak disk magnetization
and by magneto-centrifugal support in the case of strong magnetization.
Between the two parameter regimes there is a smooth transition, however the above mentioned change in slope 
indicates a {"}critical{"} disk magnetization of $\mu_{\rm D} \simeq 0.01$ that
separates the two regimes.
The existence of these two regimes has been discussed by \citet{1997A&A...319..340F} considering
the outflow ejection efficiency indices.

(7)
Our results are obtained from simulations applying a well defined magnetic diffusivity distribution for 
which the vertical profile is kept constant in time, while the strength of diffusivity changes in time
with disk magnetization. 
Our diffusivity model allows for smooth and continuous accretion for very long time scales 
\citep{2014ApJ...793...31S} and results in a disk magnetization that is about constant along the
midplane, similar to \citet{2009MNRAS.400..820T}.
When using a different scale height for the diffusivity profile we have not detected a substantial
variation in the mass loading, although we know from our previous work \citep{2012ApJ...757...65S}
that 
Insofar, different assumptions on the diffusivity profile, in particular a time evolution
in the level of diffusivity \citep{2010A&A...512A..82M,2013ApJ...774...12F} may alter our
numerical results.
Also, numerical effects may play a role \citep{2010A&A...512A..82M}, however they are difficult
to quantify.
However, while variations in the disk evolution can be expected by using a different diffusivity prescription,
we believe that our general results are indeed robust and can be applied generally.

We hesitate to speculate what kind of jet sources may have weakly or strongly magnetized disk.
The disk magnetization may depend on the source of the magnetic field (a field from a disk or stellar dynamo or 
advected from the ISM), and can be altered by diffusive, viscous and advective processes.
At this point, numerical simulations of the local disk physics are essential for our understanding.

\acknowledgements
We thank Andrea Mignone and the PLUTO team for the possibility to use their code.
We thank Rachid Ouyed for valuable comments on an earlier version of the manuscript.
We acknowledge comments and suggestions by anonymous referees.
Our simulations were performed on the THEO cluster of Max Planck Institute for Astronomy.
This work was partly financed by the SFB 881 of the German science foundation DFG.



\bibliographystyle{apj}



\end{document}